\documentclass[journal]{IEEEtran}
\usepackage {graphicx}
\usepackage{setspace}
\usepackage{amssymb}
\usepackage{amsfonts}
\usepackage{mathrsfs}
\usepackage{amssymb,amsmath}
\usepackage{epstopdf}
\usepackage{bm}
\usepackage{algorithm}
\usepackage{algorithmic}
\usepackage{amsmath,amssymb,epsfig,graphics,subfigure}
\usepackage{theorem}
\usepackage{array,color}
\usepackage[compress]{cite}
\usepackage{stfloats}
\usepackage{graphicx}
\usepackage{subfigure}
\usepackage{flushend}
\usepackage{array}
\usepackage{cases}
\usepackage{multirow}
\usepackage{chngpage}
\usepackage{longtable}
\usepackage{lineno}
\usepackage{amsfonts}
\usepackage{mathrsfs}
\usepackage{amssymb,amsmath}
\usepackage{algorithm}
\usepackage{algorithmic}
\usepackage{amsmath,amssymb,epsfig,graphics,subfigure}
\usepackage{theorem}
\usepackage{array,color}
\usepackage{amssymb}
\newcommand{\tb}{\mathbf}

\newcommand{\wj}{\textcolor{black}}
\newcommand{\rmsg}{\overrightarrow}
\newcommand{\lmsg}{\overleftarrow}

\theoremheaderfont{\normalfont\bfseries}
\begin{document}
\title{Bayesian Predictive Beamforming for Vehicular Networks: A Low-overhead Joint Radar-Communication Approach}
\author{Weijie Yuan,~\IEEEmembership{Member,~IEEE}, Fan Liu,~\IEEEmembership{Member,~IEEE}, Christos Masouros,~\IEEEmembership{Senior Member,~IEEE,} \\Jinhong Yuan,~\IEEEmembership{Fellow,~IEEE,} Derrick Wing Kwan Ng,~\IEEEmembership{Senior Member,~IEEE,} and Nuria Gonz\'{a}lez-Prelcic,~\IEEEmembership{Senior Member,~IEEE,}
\thanks{W. Yuan, J. Yuan, and D. W. K. Ng are with the School of Electrical Engineering and Telecommunications, University of New South Wales, Sydney, NSW 2052, Australia (e-mail: weijie.yuan@unsw.edu.au, j.yuan@unsw.edu.au, w.k.ng@unsw.edu.au).}
\thanks{F. Liu and C. Masouros are with the Department of Electronic and Electrical Engineering, University College London, London, WC1E 7JE, UK (e-mail: fan.liu@ucl.ac.uk, chris.masouros@ieee.org).}
\thanks{N. Gonz\'{a}lez-Prelcic is with the Department of Electrical and Computer Engineering, The University of Texas at Austin, Austin, TX 78712-1687, USA, and also with the Signal Theory and Communications Department, University of Vigo, 36310 Vigo, Spain (e-mail: ngprelcic@utexas.edu).}
}
\maketitle
\begin{abstract}
The development of dual-functional radar-communication (DFRC) systems, where vehicle localization and tracking can be combined with vehicular communication, will lead to more efficient future vehicular networks. In this paper, we develop a predictive beamforming scheme in the context of DFRC systems. We consider a system model where the road-side units estimates and predicts the motion parameters of vehicles based on the echoes of the DFRC signal. Compared to the conventional feedback-based beam tracking approaches, the proposed method can reduce the signaling overhead and improve the accuracy. To accurately estimate the motion parameters of vehicles in real-time, we 
 propose a novel message passing algorithm based on factor graph, which yields near optimal solution to the maximum \emph{a posteriori} estimation. \wj{The beamformers are then designed based on the predicted angles for establishing the communication links.} With the employment of appropriate approximations, all messages on the factor graph can be derived in a closed-form, thus reduce the complexity. \wj{Simulation results show that the proposed DFRC based beamforming scheme is superior to the feedback-based approach in terms of both estimation and communication performance.} Moreover, the proposed message passing algorithm achieves a similar performance of the high-complexity particle-based methods.
\end{abstract}


\section{Introduction}
Connectivity and automation are revolutionizing the automotive industry. These technologies are enabling innovations that will transform vehicles into platforms for drivers and passengers that will provide services beyond transportation\cite{siegel2017survey,lu2014connected}
In the era of the fifth-generation (5G) communication, vehicle-to-everything (V2X) communication is expected to play an important role to support promising applications including automated vehicles, traffic management, and social driving \cite{chen2017vehicle}. To fulfill the stringent quality of service (QoS) requirement in 5G, the V2X network is required to support low latency information transmission in high-mobility environments\cite{wang2017overview}
. In addition to the wireless communication, the environment sensing capability is also of great importance in vehicular networks
. Since its birth, radar systems have been already deployed worldwide to address several military and civilian applications such as obstacle detection, environment reconstruction, as well as remote sensing \cite{skolnik2001radar}. In the future V2X systems, due to the time-varying nature of the network topology and the surrounding environments, radar-type technologies are envisioned as a promising candidate for detecting and tracking cars, pedestrians, road lanes, and obstacles in real-time \cite{dickmann2016automotive}.

Traditionally, radar and communication systems exploit separate spectrum resources and thus rarely interfere with each other. However, this approach becomes challenging, or even infeasible, as the communities developing both types of systems are seeking more resources. 
As a result, for V2X applications, where both radar and communication functionalities are desirable, mutual interference is unavoidable in overlapped frequency bands. A straightforward solution is to allocate two systems with separated portions of frequency spectrum to eliminate any possible interference, yet, resulting in a severe degradation of the spectral efficiency. Departing from this naive approach, a recent stream of research focuses on the design of joint systems that perform both radar sensing and communication functionalities with a single transmission. That is the dual-functional radar communication (DFRC) technique, which is capable of significantly reducing the hardware cost while increasing the overall system throughput\cite{liu2020joint}. 

Early contributions of DFRC designs focused on the employment of orthogonal frequency division multiplexing (OFDM) technique\cite{sen2010adaptive}. OFDM offers several advantages for wireless communication systems, such as resilience to multi-path fading, simple time and frame synchronization, and low-complexity equalization. From the radar sensing perspective, OFDM-like radar signals decouple the range and Doppler estimators in contrast to conventional radar systems, which provides a better target estimation performance \cite{shi2017power}. Thus, OFDM has been regarded a promising candidate for implementing DFRC systems \cite{sturm2011waveform,chiriyath2017radar}. 
To satisfy the requirement of communication with multiple users, DFRC systems based on the multiple-input multiple-output (MIMO) technology have been proposed in recent work. In MIMO radar, the waveforms radiated by each antenna element are orthogonal, thus the interference can be avoided and more degrees of freedom are available for DFRC waveform design. Pioneered by \cite{kumari2015investigating}
, a MIMO DFRC approach is proposed to detect target using the mainlobe of the spatial beampattern, while transmitting useful information by varying the sidelobe power. Later on, more advanced DFRC approaches were studied under various conditions and constraints, such as Non Line-of-Sight (NLoS) channels \cite{liu2018mu}, multi-user interference and peak-to-average-power ratio (PAPR)-constrained transmission \cite{liu2018toward}. However, the aforementioned works focus on the sub-6G Hz frequency band and cannot support Gbps data rate as required by V2X communication systems.

To overcome the above challenges, the large bandwidth available in the millimeter wave (mmWave) spectrum is considered as a key enabler for DFRC systems, which can significantly improve both the data rate for communication and the range resolution for radar. In particular, by leveraging the massive MIMO technology, mmWave systems are able to exploit  efficient spatial processing methods such as beamforming and spatial multiplexing at the transmitter and/ or receiver sides \cite{sun2014mimo,venugopal2016device}. Aiming for designing DFRC transceivers at the mmWave band, the authors of \cite{kumari2019adaptive} designed an adaptive DFRC waveform that meets the Pareto optimal bound. Then a novel framework based on hybrid analog-digital beamforming techniques was developed in \cite{liu2020joint}. However, \cite{liu2020joint} did not take the high-mobility environments into account and their results are thus not suitable for V2X applications. 
Relying on the large-scale antenna array, the pencil-like spatial beams can be generated by the transmitter focusing the radiation power on the intended directions, which compensates for the high path-loss of the mmWave signals. To establish a reliable communication link, it is essential to align the transmit and receive beams between the vehicles and the associated road-side unit (RSU) \cite{haghighatshoar2016beam}. Conventionally, the RSU periodically scans the angular interval of interest and pairs the transmit and receive beams with the strongest channel gain. However, for the vehicular applications with dynamic network topology and environments, the beam pairing has to be done frequently, leading to exceedingly high communication cost.

To cope with the high-mobility constraint imposed on the V2X scenarios, several works considered the fast beam tracking problem from the communication perspective\cite{va2016beam,zhang2019codebook}.
In \cite{zhang2019codebook}, the RSU first sends a communication signal containing pilots to the vehicles. Then the vehicles decode the information and estimate the relative angles with respect to the RSU, which are then fedback to the RSU for beam steering to the intended direction. To reduce the latency for beam alignment, the RSU can have a further prediction of the relative angles in the next time slot. Then the RSU can adopt their beamformers ahead for establishing the communication link in the following time instant. Some recent works utilized the classic Extended Kalman filtering (EKF) based on the state evolution of vehicles to predict the angles \cite{shaham2019fast}. Except for the angular parameters, it is important to track the variation of other kinetic motion parameters of vehicles such as speed and range to fulfill the requirements of intelligent traffic applications. To achieve highly accurate estimation result, the number of pilots for EKF beam tracking should be sufficiently large, leading to high communication signaling overhead. Some radar aided beam alignment method were proposed in \cite{gonzalez2016radar,ali2019millimeter},
where the radar signal operates in a different frequency band, which however leads to increased spectral consumption. For these reasons, we develop a DFRC-based scheme for tracking the beam direction as well as the motion parameters in vehicular networks.

In this paper, we propose a novel DFRC-based predictive beamforming scheme for vehicle-to-infrastructure (V2I) scenarios that has a very low signaling overhead. To elaborate, the RSU sends DFRC signals containing information to the vehicles via downlink transmission. After receiving the DFRC signals, the vehicles perform detection followed by decoding to obtain the information. Meanwhile, the echoes reflected by the vehicles are acquired by the RSU via radar sensing techniques to estimate the beam directions as well as other motion parameters of the vehicles. With the help of the radar system, feedbacks from the vehicles to the RSU can be avoided. Compared to some feedback-based schemes that exploits a limited number of pilots for beam tracking, e.g. \cite{shaham2019fast}, the proposed DFRC-based scheme utilizes the whole downlink block both as communication data symbols and sensing pilots. Therefore the signaling overhead solely for beam alignment is as low as $0$, which undoubtedly increases the spectral efficiency. Accordingly, uplink resources can be fully exploited for transmitting data rather than carrying feedback information. Different from the feedback-based methods, the ``prediction'' is done at the RSU, and the predicted angular parameters are contained into the DFRC signals sent to the vehicles. To fully realize the DFRC-based predictive beamforming, we formulate the beam tracking problem as a Bayes inference. In contrast to our previous work \cite{liu2020radar}, we commence from the optimal estimation perspective and  propose a factor graph and message passing-based algorithm to estimate and predict the motion parameters of vehicles at each time slot. Compared to the EKF method in \cite{liu2020radar}, the proposed approach is expected to achieve a better estimation performance. Note that the product factor node and the nonlinear functions concerning the angular parameters complicate the considered problem making the standard message passing algorithm inapplicable to provide any closed-form solutions. In general, particle filtering (PF) can be used for accurately approximating the non-closed form messages. However, the employment of a large number of particles results in high complexity. To strike a balance between the complexity and performance, we propose a modified mean field (MF) message passing algorithm and apply second-order Taylor expansion to linearize the inverse trigonometry functions. As a result, all messages are derived in a closed-form, leading to a low level of  computational complexity. In a nutshell, our contributions are summarized as follows:
\begin{itemize}
  \item We propose a DFRC based predictive beamforming scheme for vehicular networks, achieving a better estimation performance and higher spectral efficiency than that of conventional feedback-based schemes. \wj{We consider a prediction of the angles for formulating the beamformers in order to reduce the latency.}
  \item We introduce a specifically tailored factor graph-based framework and a message passing algorithm to accurately track and predict the motion parameters of vehicles.
  \item We exploit the MF message passing and Taylor expansion approaches for deriving closed form messages on factor graph, providing a low-complexity and high-accuracy estimation result.
\end{itemize}
Simulation results show that our proposed algorithm achieves a similar performance of the PF-based method but with only a considerably lower computational overhead. Moreover, compared to the communication-only feedback scheme \cite{shaham2019fast}, the proposed algorithm achieves better tracking performance and higher achievable rate.

The remainder of this paper is organized as follows. We introduce the DFRC signal model and the kinematic model of vehicles in Section II. Section III constructs the factor graph based on the probabilistic model of the considered system. In Section IV, the proposed message passing algorithm for beam tracking and motion parameter estimation is described. Then the simulation results are shown and discussed in Section V. Finally, our conclusions are drawn in Section VI.

	\emph{Notations:} We use a boldface letter to denote a vector. The superscripts $(\cdot)^{\rm T}$, $(\cdot)^{-1}$ {and} $(\cdot)^{\rm H}$ denote the transpose, the conjugate, {the} inverse and the Hermitian operations{, respectively}; $\mathcal{N}(x;{m}_{x},{\lambda}_{x})$ denotes the Gaussian distribution of real variable ${x}$ having mean of ${m}_{x}$ and variance of ${\lambda}_{x}$; $\mathcal{S}\backslash x$ denotes all variables in set $\mathcal{S}$ except $x$; $\mathbb{E}$ represents the expectation operator; $\mathbb{V}$ denotes the operator to obtain the variance of a random variable; $\mathbb{C}^{N\times M}$ denotes a complex space of dimensional $N\times M$; $\mathcal{R}\{\cdot\}$ and $\mathcal{I}\{\cdot\}$ denote the real part and imaginary part of a complex variable; $|\cdot|$ represents the modulus of a complex number; $\propto$ represents both sides of the equation are multiplicatively connected to a constant; $\textrm{erf}(\cdot)$denotes the error function; $\tb{x}\backslash x$ denotes all variables in $\tb{x}$ except $x$.

\section{System Model}
\begin{figure}[h]
	\centering
	\includegraphics[width=.4\textwidth]{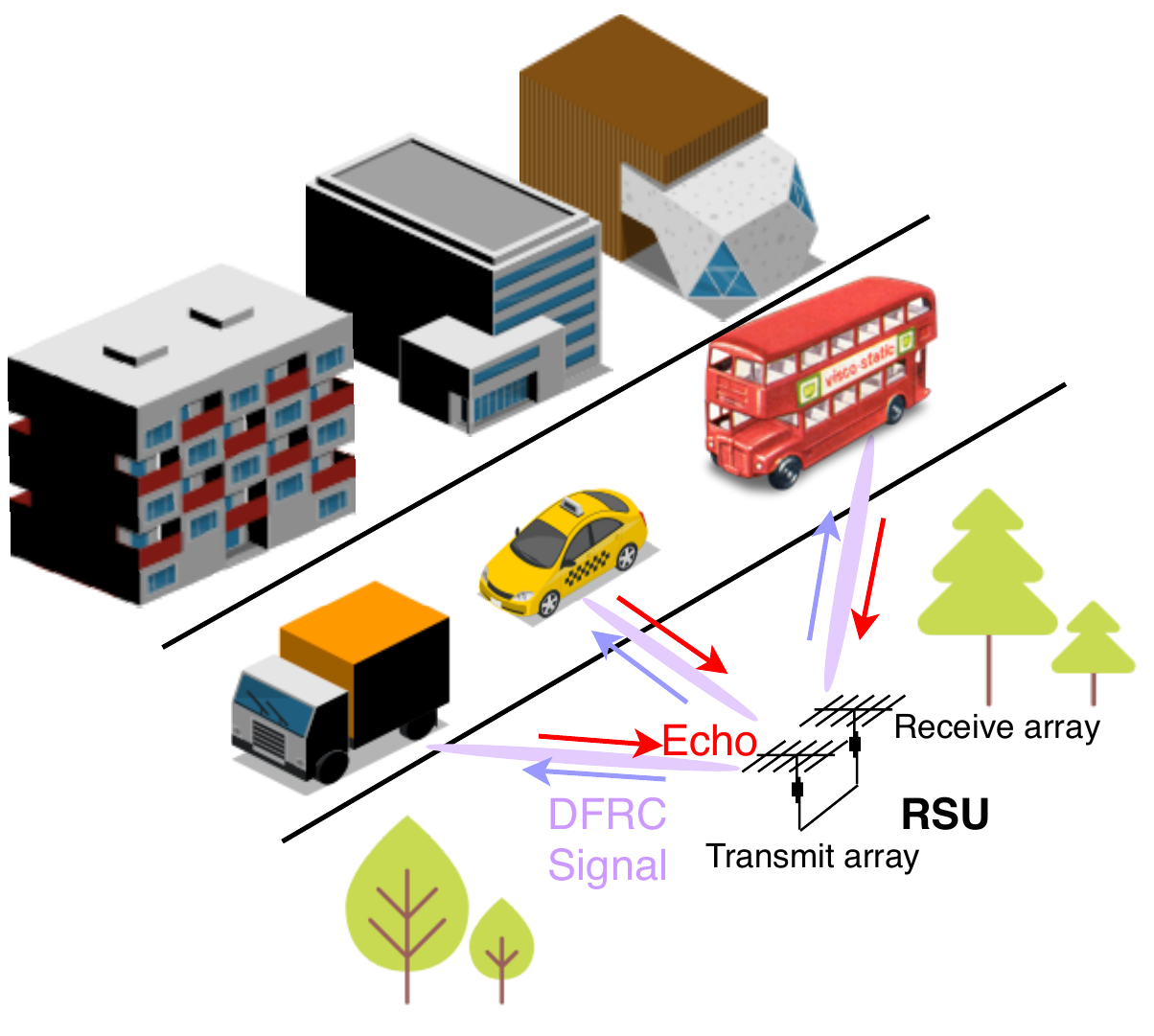}
	\caption{System model for the considered vehicular network.}\label{model1}
	\centering
\end{figure}

Throughout this paper, we consider a vehicular network with one RSU supporting $K$ vehicles, as depicted in Fig. \ref{model1}. The RSU operates at mmWave band equipped with a massive MIMO uniform linear array (ULA) which has $N_t$ transmit antennas and a separate array of $N_r$ receive antennas. This allows the RSU to receive the vehicle echoes for tracking while ensuring uninterrupted downlink transmission. Each one of the vehicles is assumed to have an $M$-antenna ULA mounted on both sides of its body. The vehicles move along a one-lane straight road parallel to the RSU array\footnote{For convenience of exposition, the vehicles are modeled as point targets following standard assumption in the literature \cite{wymeersch20175g}. The extension to a non-parallel case is straightforward since the angle between the road and the RSU array is fixed and known to the RSU and can be easily calibrated.}. \wj{For brevity, we assume that the signals transmit through line-of-sight (LoS) channels, a further extension to non line-of-sight channels will be designated to the future works.}

Without loss of generality, we denote the range, the angle, and the speed of the $k$th vehicle relative to the RSU's array by $d_{k}(t)$, $\theta_k(t)$, and $v_k(t)$, respectively. Provided that the vehicle is moving along the direction parallel with the ULA, the angle of the RSU relative to the $k$th vehicle can also be defined as $\theta_k(t)$. We further denote the time duration between two radar epochs by $T$. Following commonly adopted assumptions\cite{shaham2019fast}, we assume that the required parameters for vehicle $k$ do not change in a relatively short time duration. Then the motion parameters of vehicle $k$ at the $n$th time instant are denoted by $d_{k,n}$, $\theta_{k,n}$, and $v_{k,n}$.
\subsection{Signal Model}

At the $n$th instant, the RSU sends a $K$-dimensional multi-beam direction-finding DFRC signal to the $K$ vehicles concurrently, denoted by $\tb{s}_n(t) = [s_{1,n}(t),...,s_{K,n}(t)]^{\rm T}$ with a complex signal $s_{k,n}(t)$ for vehicle $k$. The signal $\tb{s}_n(t)$ is transmitted over ${N}_t$ antennas of the RSU, formulating as
\begin{align}
\tilde{\tb{s}}_n(t) =\tb{F}_n\tb{s}_n(t),
\end{align}
where $\tb{F}_n\in\mathbb{C}^{N_t\times K}$ is the transmit beamforming matrix. In general, the beamforming matrix $\tb{F}_n$ is designed relying on the predicted angle. Assuming that we have a prediction of angle $\theta_k$ for all vehicles $\forall k\in [1,K]$, defined by ${\theta}^{\rm pred, R}_{k,n}$, the beamforming vector for the $k$th vehicle is the $k$th column of $\tb{F}_n$ that can be written as
\begin{align}\label{bf_vector}
\tb{f}_{k,n} = \sqrt{e_{k,n}}\tb{a}({\theta}^{\rm pred, R}_{k,n}),
\end{align}
where $e_{k,n}$ denotes the signal power, $\tb{a}({\theta}^{\rm pred, R}_{k,n})=\sqrt{\frac{1}{N_t}}[a_1({\theta}^{\rm pred, R}_{k,n}),...,a_{N_t}({\theta}^{\rm pred, R}_{k,n})]^{\textrm{T}}$ with $a_i({\theta}) =e^{-j\pi (i-1)\cos {\theta}}$ is the transmit steering vector and ${\theta}^{\rm pred, R}_{k,n}$ is the predicted angle at time $n$. By using the beamforming matrix $\tb{F}_n$, each beam is steered towards the predicted angle of the targeted vehicle. The transmitted signal $\tilde{\tb{s}}_n(t)$ is reflected by all the $K$ vehicles and the received echo $\tb{r}_n(t) = [r_{1,n}(t),...,r_{K,n}(t)]^{\rm T}$can be formulated as
\begin{align}
\tb{r}_n(t) =& \varsigma \sum_{k=1}^K \beta_{k,n}e^{j2\pi \gamma_{k,n}t} \tb{b}(\theta_{k,n}) \tb{a}^{\rm H} (\theta_{k,n})\tilde{\tb{s}}_{n}(t-\tau_{k,n}) \nonumber\\&+ \tb{z}_r(t),
\end{align}
where $\varsigma=\sqrt{N_t N_r}$ denotes the multi-antenna array gain, $\beta_{k,n}$, $\gamma_{k,n}$, and $\tau_{k,n}$ denote the reflection coefficient, the Doppler, and the delay of the $k$th vehicle at time $n$, respectively, and $\tb{b}(\theta_{k,n})=\sqrt{\frac{1}{N_r}}[b_1(\theta_{k,n}),...,b_{N_r}(\theta_{k,n})]^{\textrm{T}}$ is the receive steering vector with $b_i(\theta) =e^{-j\pi (i-1)\cos \theta}$\footnote{It should be noted that we are considering a fully digital massive MIMO architecture, where the echo signal (3) is received by the RSU without analog combination.}. The term ${\tb{z}}_r(t)$ is assumed to be a complex additive white Gaussian noise with zero mean.

{\emph{Remark 1:} The echoes may arrive at the RSU during the transmission of the DFRC signal, in which case the transmitted signal would inevitably generate self-interference to the reception. To address this issue, we employ separate transmit and receive antenna arrays at the RSU, with strong RF isolation combined with highly directive elevation beamforming. Then, the self-interference can be cancelled out via exploiting classic RF full-duplex structure \cite{palacios2019hybrid}, where the residual interference is small enough to be incorporated into the noise term. For notational convenience and to avoid alleviating from the focus of this paper, we will not discuss this topic in detail and designate this as our future work. \hfill $\blacksquare$}

Given the relative range of vehicle $k$ and the RSU, i.e., $d_{k,n}$, the reflection coefficient can be modeled as
\begin{align}\label{beta_model}
\beta_{k,n}=\frac{\xi}{2{d_{k,n}}},
\end{align}
where $\xi$ represents the complex radar cross-section (RCS)\cite{skolnik2001radar}.

According to \cite{marzetta2016fundamentals}, the following result generally holds for ULA in a massive MIMO scenario:
\begin{align}
|\tb{a}^{\rm H}(\theta_i)\tb{a}(\theta_j)|\approx 0,~\forall i\neq j.
\end{align}
This result indicates that the steering vectors related to two different vehicles are asymptotically orthogonal in massive MIMO systems \cite{shaham2019fast}. Therefore, we assume that there is only negligible inter-beam interference and the RSU can identify the echoes from different vehicles. The reflected echo for the $k$th vehicle can then be expressed as
\begin{align}\label{rx_echo}
\tb{r}_{k,n}(t) = &\varsigma \beta_{k,n}e^{j2\pi \gamma_{k,n}t} \tb{b}(\theta_{k,n}) \tb{a}^{\rm H} (\theta_{k,n})\tilde{\tb{s}}_{k,n}(t-\tau_{k,n}) \nonumber\\&+ \tb{z}_{k,n}(t),
\end{align}
where $\tilde{\tb{s}}_{k,n}(t) =\tb{f}_{k,n}{{s}}_{k,n}(t)$. In the following section, we will introduce the measurement model adopted at the RSU.

\subsection{Observation Model}
By performing radar matched filtering on \eqref{rx_echo} with a delayed and Doppler-shifted version of $\tb{s}_{k,n}(t)$, we obtain the estimates of  delay $\tau_{k,n}$ and Doppler $\gamma_{k,n}$, which are related to range $d_{k,n}$ and speed $v_{k,n}$, respectively. Considering Gaussian observation noise, the measurement model for the motion parameters $d_{k,n}$ and $v_{k,n}$ are given by
\begin{align}\label{delay}
\tau_{k,n} &= \frac{2d_{k,n}}{c} + z_\tau,~\textrm{and}\\\label{doppler}
\gamma_{k,n} &= \frac{2v_{k,n} \cos \theta_{k,n} f_c}{c} + z_\gamma,
\end{align}
respectively, where $f_c$ and $c$ represent the carrier frequency and signal propagation speed, noise terms $z_\tau$ and $z_\gamma$ obey Gaussian distributions $\mathcal{N}(z_\tau;0,\sigma_\tau^2)$ and $\mathcal{N}(z_\gamma;0,\sigma_\gamma^2)$, respectively.

\emph{Remark 2:} By only exploiting the observation within a single time-slot, the speed estimation of a vehicle at the RSU is challenging, since the resultant Doppler phase shift is not significant \cite{eaves2012principles}. To tackle this issue, one may estimate the vehicle's velocity given a Doppler shift accumulated within multiple time-slots. \hfill $\blacksquare$

Having the estimates of delay and Doppler, we have the received signal samples ${\tb{y}}_{k,n}=[y_{k,n}^{[1]},...,y_{k,n}^{[N_r]}]^T$ for $\theta_{k,n}$ and $\beta_{k,n}$ based on the filtered signal, given by
\begin{align}\label{observation_y}
{\tb{y}}_{k,n} = \varsigma\beta_{k,n}\sqrt{e_{k,n}} \tb{b}({\theta}_{k,n})\tb{a}^\textrm{H}({\theta}_{k,n}) \tb{a}({\theta}^{\rm pred, R}_{k,n}) +{\tb{z}}_{k,n},
\end{align}
where the term $\tb{z}_{k,n} = [z_{k,n}^{[1]},...,z_{k,n}^{[N_r]}]^{\rm T}$ denotes the noise samples at different receive antennas. Without loss of generality, we model ${z}^{[i]}_{k,n}=z_y \sim\mathcal{N}(z_y;0,\sigma_y^2)$,~$\forall i$. Remark that after matched filtering, we achieve a signal-to-noise ratio (SNR) gain $G$, which is typically identical to the energy of signal $s_{k,n}(t)$. Vector $\tb{a}({\theta}^{\rm pred, R}_{k,n})$ in \eqref{observation_y} denotes the transmit beamformer for vehicle $k$ based on the predicted angle ${\theta}^{\rm pred, R}_{k,n}$ at time instant $n$. After straightforward manipulations, we arrive at the following measurement model
\begin{align}\label{model}
{\tb{y}}_{k,n} = &\beta_{k,n} p_{k,n} \left[
\begin{array}{c}
\sum\limits_{i=1}^{N_t}e^{j\pi(i-1)\cos\hat{\theta}_{k,n}} e^{-j\pi(i-1)\cos\theta_{k,n}}\\
\sum\limits_{i=1}^{N_t}e^{j\pi(i-1)\cos\hat{\theta}_{k,n}} e^{-j\pi(i-2)\cos\theta_{k,n}}\\
...\\
\sum\limits_{i=1}^{N_t}e^{j\pi(i-1)\cos\hat{\theta}_{k,n}} e^{-j\pi(i-N_r)\cos\theta_{k,n}}
\end{array}
\right]\nonumber\\&+{\tb{z}}_{y},
\end{align}
where $p_{k,n}=\sqrt{\frac{e_{k,n}}{N_t}}$. Next, we consider the state evolution model of the motion parameters of vehicles.

\subsection{State Evolution Model}
 \begin{figure}[]
	\centering
	\includegraphics[width=.5\textwidth]{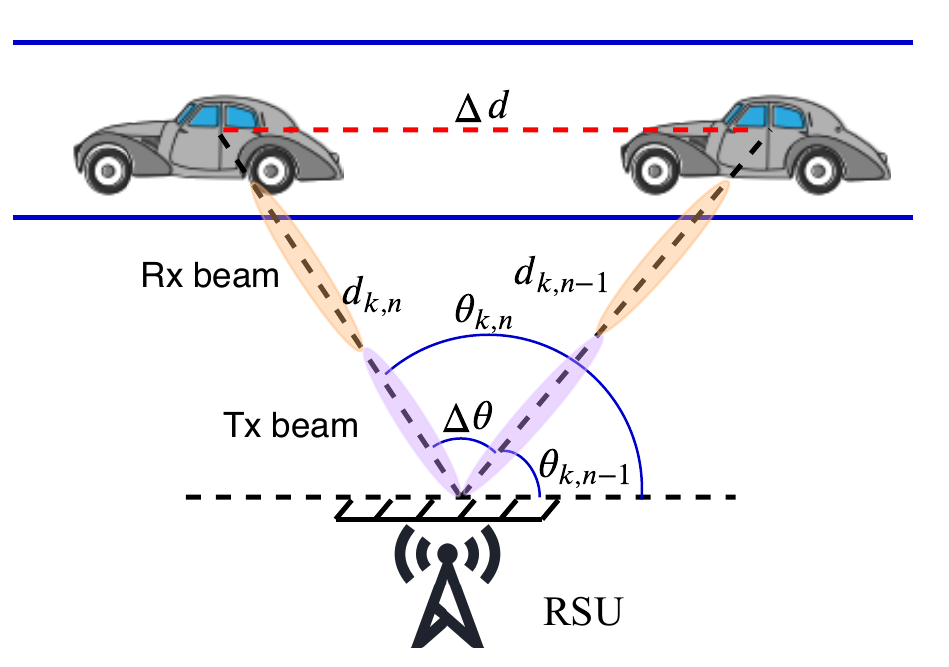}
	\caption{State evolution model of the considered vehicular network.}\label{model2}
	\centering
\end{figure}

We aim for obtaining the motion parameters of all vehicles in the coverage area of the RSU. Based on the previous states and the moving patterns of the vehicles, we can determine the state evolution models for the vehicles, as shown in Fig. \ref{model2}. Relying on the geometric relationship of the motion parameters at time instant $n-1$ and $n$ as shown in Fig. \ref{model2}, we have the following kinematic equations as
\begin{align}\label{kinematic}
\left\{\begin{array}{l}
\sin (\theta_{k,n}-\theta_{k,n-1})d_{k,n} = v_{k,n-1} T \sin \theta_{k,n-1},\\
d_{k,n}^2 =  d_{k,n-1}^2 + (v_{k,n-1} T)^2 - 2d_{k,n-1}v_{k,n-1} T\cos \theta_{k,n-1}.
\end{array}
\right.
\end{align}
The above two equations show how the motion parameters of vehicle $k$ evolve with time. For brevity, we define $\Delta \theta = \theta_{k,n}-\theta_{k,n-1}$ and $\Delta d = v_{k,n-1} T$. Obviously, solving the above nonlinear equations to construct the evolution model is challenging. As a compromise approach, we propose to find a tractable approximation of \eqref{kinematic}. Since in general, the variation of vehicle's position is relatively small within a short period $T$, we can use the approximation $\Delta \theta_{k,n} \approx \sin\Delta \theta_{k,n}$. Based on this, the first equation of \eqref{kinematic} can be written as
\begin{align}
\Delta \theta_{k,n} \approx \sin\Delta \theta_{k,n} = \frac{\Delta d\sin \theta_{k,n-1}}{d_{k,n}}.
\end{align}
By considering the fact that $d_{k,n}\approx d_{k,n-1}$ in time duration $T$, $\Delta \theta_{k,n}$ is further expressed as
\begin{align}\label{theta_model}
\Delta \theta_{k,n}\approx \frac{\Delta d\sin \theta_{k,n-1}}{d_{k,n-1}}.
\end{align}
For the range variation in two continuous time slots, we can rewrite the second equation of \eqref{kinematic} as
\begin{align}\label{variation_d}
d_{k,n}-d_{k,n-1}=\frac{\Delta d^2 -2d_{k,n-1}\Delta d\cos \theta_{k,n-1}}{d_{k,n}+d_{k,n-1}}.
\end{align}
Then, by exploiting the fact that $d_{k,n}\approx d_{k,n-1}$, \eqref{variation_d} can be simplified to
\begin{align}\label{variation_d1}
d_{k,n}-d_{k,n-1}\approx \Delta d\left(\frac{\Delta d}{2d_{k,n-1}}-\cos \theta_{k,n-1}\right).
\end{align}
Comparing to the range $d_{k,n-1}$, the variation $\Delta d$ is negligible, therefore the term $\frac{\Delta d}{2d_{k,n-1}}$ can be omitted and $d_{k,n}-d_{k,n-1}$ is given by
\begin{align}\label{variation_d2}
d_{k,n}-d_{k,n-1}\approx \Delta d \cos \theta_{k,n-1}.
\end{align}
Furthermore, using the reflection coefficient model \eqref{beta_model}, we have
\begin{align}\label{beta_model_2}
\beta_{k,n} &= \beta_{k,n-1}\frac{\xi d_{k,n-1}}{\xi d_{k,n}}\nonumber\\
&= \beta_{k,n-1}\approx\left(1+\frac{\Delta d \cos\theta_{k,n-1}}{d_{k,n-1}}\right).
\end{align}
Finally, we assume that the speed of the vehicle is approximately a constant, i.e.,
\begin{align}\label{v_model}
v_{k,n}\approx v_{k,n-1}.
\end{align}

Having \eqref{theta_model}-\eqref{v_model}, we summarize the state evolution models for $\theta_{k,n}$, range $d_{k,n}$, speed $v^{k,n}$, and coefficient $\beta_{k,n}$ in the following
\begin{align}\label{state_evolution_theta}
\theta_{k,n} &= \theta_{k,n-1}+\frac{v_{k,n-1} T \sin \theta_{k,n-1}}{d_{k,n-1}}+z_\theta,\\\label{state_evolution_d}
d_{k,n} &= d_{k,n-1}-v_{k,n-1} T\cos \theta_{k,n-1}+ z_d,\\\label{state_evolution_v}
v_{k,n} &= v_{k,n-1}+z_v,\\\label{state_evolution_beta}
\beta_{k,n} &= \beta_{k,n-1}+\beta_{k,n-1}\frac{v_{k,n-1} T \cos \theta_{k,n-1}}{d_{k,n-1}}+z_\beta,
\end{align}
where the transition noise $z_\theta$, $z_d$, $z_v$, and $z_\beta$ obey zero mean Gaussian distributions $\mathcal{N}(z_\theta;0,\sigma_\theta^2)$, $\mathcal{N}(z_d;0,\sigma_d^2)$, $\mathcal{N}(z_v; 0,\sigma_v^2)$, and $\mathcal{CN}(z_\beta; 0,\sigma_\beta^2)$, respectively. Note that the state evolution of $d_{k,n}$, $\theta_{k,n}$, and $\beta_{k,n}$ also depend on other variables. For simplicity, we adopt the estimates at time instant $n-1$, i.e., $\hat{v}_{k,n-1}$, $\hat{\theta}_{k,n-1}$, and $\hat{d}_{k,n-1}$ to replace the corresponding terms in \eqref{state_evolution_theta}-\eqref{state_evolution_beta}, such that the evolution of the motion parameters only depends on their own previous states.

\subsection{Communication Model}
To receive the signal sent by the RSU, vehicle $k$ adopts a receive beamformer $\tb{w}_{k,n}$ and the received signal is formulated as
\begin{align}
g_{k,n}(t) = \bar{\varsigma}\alpha_{k,n}\tb{w}^H_{k,n} \tb{u}(\theta_{k,n})\tb{a}^{\rm H}(\theta_{k,n})\tilde{\tb{s}}_{k,n}(t) + {z}_g(t),
\end{align}
where $\bar{\varsigma}=\sqrt{N_t M}$ is the array gain between the RSU and the vehicle, $\alpha_{k,n}$ is the channel pathloss coefficient, ${z}_g(t)$ is the Gaussian noise term, and $\tb{u}(\theta_{k,n}) \in \mathbb{C}^{M\times 1}$ denotes the receive steering vector of vehicle $k$, which has a similar definition as $\tb{a}(\theta)$. The beamformer $\tb{w}_{k,n}$ is designed based on the predicted angle of vehicle $k$ relative to the RSU at time instant $n$, i.e., $\tb{w}_{k,n} = \tb{u}({\theta}^{\rm pred}_{k,n})$. The prediction of the angle ${\theta}^{\rm pred}_{k,n}$ at time $n$ is done at the RSU based on the estimate at time $n-2$, which is contained in the DFRC signal at time $n-1$ to the $k$th vehicle. Assuming that the original transmitted signal ${s}_{k,n}(t)$ from the RSU has unit power, then the SNR of the received signal is given by
\begin{align}\label{SNR_exp}
\textrm{SNR}_{k,n} = \frac{\left|\bar{\varsigma}\alpha_{k,n}\tb{w}^H_{k,n} \tb{u}(\theta_{k,n})\tb{a}^{\rm H}(\theta_{k,n})\tb{f}_{k,n}\right|^2}{N_0},
\end{align}
where $N_0$ is the power spectral density (PSD) of the additive white Gaussin noise.
Substituting the expression of $\tb{f}_{k,n}$ into \eqref{SNR_exp} yields
\begin{align}
\textrm{SNR}_{k,n} = \bar{\varsigma}^2 e_{k,n} \frac{\left|\alpha_{k,n}\tb{u}^{\rm H}({\theta}^{\rm pred}_{k,n}) \tb{u}(\theta_{k,n})\tb{a}^H(\theta_{k,n})\tb{a}(\hat{\theta}_{k,n|n-1})\right|^2}{N_0}.
\end{align}
Based on the SNR corresponding to the $k$th vehicle, the achievable sum-rate of all vehicles at time $n$ is expressed as
\begin{align}
  R_n = \sum_{k=1}^K (1+\textrm{SNR}_{k,n}).
\end{align}
It can be observed that the achievable sum-rate relies on the transmit and receive beamformers. When the angle is perfectly predicted, i.e., $\theta_{k,n} = {\theta}^{\rm pred,R}_{k,n} = {\theta}^{\rm pred}_{k,n}$, $\tb{u}^{\rm H}({\theta}^{\rm pred}_{k,n}) \tb{u}(\theta_{k,n})=1$ and $\tb{a}^H(\theta_{k,n})\tb{a}(\hat{\theta}_{k,n|n-1})$, the received SNR is maximized, given by
\begin{align}
\textrm{SNR}_{k,n} = \frac{e_{k,n}|\alpha_{k,n}|^2}{N_0},
\end{align}
and we have the maximum achievable sum-rate. For the channel coefficient $\alpha_{k,n}$, it can be simply estimated based on the range parameter $d_{k,n}$.

For clarity, we define vectors ${\tb{y}}_k = [{\tb{y}}^{\rm T}_{k,1},...,{\tb{y}}^{\rm T}_{k,N}]^{\rm T}$, $\bm{\tau}_k=[\tau_{k,1},...,\tau_{k,N}]^{\rm T}$, and $\bm{\gamma}_k = [\gamma_{k,1},...,\gamma_{k,N}]^{\rm T}$ as the received signals, observed delays and Dopplers of vehicle $k$ until time instant $N$, respectively. Furthermore, the unknown parameters corresponding to vehicle $k$ can also be rewritten in vector form as $\bm{\theta}_k = [\theta_{k,1},...,\theta_{k,N}]^{\rm T}$, $\tb{d}_k=[d_{k,1},...,d_{k,N}]^{\rm T}$, $\tb{v}_k=[v_{k,1},...,v_{k,N}]^{\rm T}$ and $\bm{\beta}_k = [\beta_{k,1},...,\beta_{k,N}]^{\rm T}$, respectively. As the echo signals reflected by different vehicles can be identified unambiguously, in what follows, we will omit the vehicle index `$k$' for brevity. Note that the proposed algorithm is applicable for any vehicles in the RSU's coverage area. In the above, we have expressed the radar signal as well as the communication model. In the next section, we will formulate a factor graph model to infer the variables representing the motion parameters of vehicles.

\section{Factor Graph Model}
Our goal is to estimate the unknown range $d$, speed $v$, angle $\theta$, and path loss $\beta$ given the state evolution model and the observation model. From the Bayesian perspective, this is equivalent to inferring the variables from the observations through an estimator, i.e., the maximum \emph{a posteriori} (MAP) estimator,
\begin{align}\label{MAP_est}
\{\hat{\tb{d}},\hat{\bm{\theta}},\hat{\tb{v}},\hat{\bm{\beta}}\} = \arg\max_{\tb{d},\bm{\theta},\tb{v},\bm{\beta}}p(\tb{d},\bm{\theta},\tb{v},\bm{\beta}|\tb{y},\bm{\tau}, \bm{\gamma}),
\end{align}
where $p(\tb{d},\bm{\theta},\tb{v},\bm{\beta}|\tb{y},\bm{\tau}, \bm{\gamma})$ denotes the joint \emph{a posteriori} distribution. Nevertheless, solving \eqref{MAP_est} involves a multi-dimensional search, leading to an exponentially increased complexity. As a result, there is a need for a low-complexity estimation approach. To this end, we aim for finding a suboptimal solution that the maximization is performed based on the marginal \emph{a posteriori} of a variable of interest $x$, formulated as $\hat{x} = \arg\max_x p(x|\tb{y},\bm{\tau}, \bm{\gamma})$. Generally, the marginal distribution can be obtained through direct marginalization of the joint distribution, i.e.,
\begin{align}\label{marginalization}
p(x|\tb{y},\bm{\tau}, \bm{\gamma}) = \int_{\{\tb{d},\bm{\theta},\tb{v},\bm{\beta}\}\backslash x}p(\tb{d},\bm{\theta},\tb{v},\bm{\beta}|\tb{y},\bm{\tau}, \bm{\gamma}).
\end{align}
However, direct marginalization requires a prohibitively high complexity due to the multi-dimensional integrations involved in \eqref{marginalization}. In what follows, we will resort to the factor graph framework to obtain the marginals of unknown variables by leveraging the conditional independency between variables.

According to Bayes Theorem, the joint distribution is rewritten as
\begin{align}\label{post_factor}
p(\tb{d},\bm{\theta},\tb{v},\bm{\beta}|\tb{y},\bm{\tau},\bm{\gamma},\bm{\beta}) = p(\tb{y},\bm{\tau}, \bm{\gamma}|\tb{d},\bm{\theta},\tb{v}) p(\tb{d},\bm{\theta},\tb{v},\bm{\beta}),
\end{align}
where $p(\tb{y},\bm{\tau}, \bm{\gamma}|\tb{d},\bm{\theta},\tb{v},\bm{\beta})$ and $p(\tb{d},\bm{\theta},\tb{v},\bm{\beta})$ are the likelihood function and the joint \emph{a priori} distribution, respectively. Let us consider the \emph{a priori} distribution first. Based on the state transition function in \eqref{state_evolution_theta}-\eqref{state_evolution_beta}, the joint \emph{a priori} distribution can be factorized as
\begin{align}\label{prior}
p(\tb{d},\bm{\theta},\tb{v},\bm{\beta})&= p(\tb{d})p(\bm{\theta})p(\tb{v})p(\bm{\beta})\nonumber\\
&=p(d_{0})p(\theta_{0})p({v}_{0})p({\beta}_{0})\prod_{n=1}^N p(d_{n}|d_{n-1})\nonumber\\&~\cdot p(\theta_{n}|\theta_{n-1})p(v_{n}|v_{n-1})p(\beta_{n}|\beta_{n-1}),
\end{align}
with the transition probabilities
\begin{align}\label{transition_1}
&p(v_{n}|v_{n-1}) \propto \exp\left(-(v_{n}-v_{n-1})^2/2\sigma_v^2\right),\\\label{transition_2}
&p(\theta_{n}|\theta_{n-1})\propto\\& \hspace{3mm}\exp\!\!\left(\!-\left(\!\theta_{n}-\theta_{n-1}-\frac{\hat{v}_{n-1}T\sin\hat{\theta}_{n-1}}{\hat{d}_{n-1}}\!\right)^2\!\!/2\sigma_\theta^2\!\right),\nonumber \\
&p(d_{n}|d_{n-1})\propto\\& \hspace{3mm}\exp\big(-(d_{n}-d_{n-1}+\hat{v}_{n-1} T\cos\hat{\theta}_{n-1})^2/2\sigma_d^2\big),\nonumber\\\label{transition_4}
&p(\beta_{n}|\beta_{n-1}) \propto \exp\left(-(\beta_{n}-\rho_{n-1} \beta_{n-1})^2/2\sigma_\beta^2\right),
\end{align}
where $\rho_{n-1}=1+\frac{\hat{v}_{k,n-1} T \cos \hat{\theta}_{k,n-1}}{\hat{d}_{k,n-1}}.$
The initial vehicle parameters at time instant $0$ are obtained by employing an \emph{omnidirectional} probing waveform sent from the RSU. Based on the received echoes, the RSU is able to infer the parameters $d_{0},~\theta_{0}$, $v_{0}$, and $\beta_0$ of a vehicle entering the coverage area of RSU. Without loss of generality, we model the initial distributions of $d_{0},~\theta_{0}$, $v_{k,0}$, and $\beta_0$ as Gaussian distributions $p(d_{0})=\mathcal{N}(d_0;m_{{d}_{0}},\lambda_{d_{0}})$, $p(\theta_{0})=\mathcal{N}(\theta_0;m_{{\theta}_{0}},\lambda_{\theta_{0}})$, $p(v_{0})=\mathcal{N}(v_0;m_{{v}_{0}},\lambda_{v_{0}})$, and $p(\beta_{0})=\mathcal{N}(\beta_0;m_{{\beta}_{0}},\lambda_{\beta_{0}})$.

For the joint likelihood function, since the received signals, observed delays and Dopplers are irrelevant given the variables, we can express the joint likelihood function as
\begin{align}
p(\tb{y},\bm{\tau}, \bm{\gamma}|\tb{d},\bm{\theta},\tb{v},\bm{\beta}) = p(\tb{y}|\bm{\theta},\bm{\beta}) p(\bm{\tau}|\tb{d}) p(\bm{\gamma}|\bm{\theta},\tb{v}).
\end{align}
Considering that the Gaussian noise terms are independent for different time instants $n$, therefore $p(\tb{y},\bm{\tau}, \bm{\gamma}|\tb{d},\bm{\theta},\tb{v},\bm{\beta})$ can be factorized as
\begin{align}\label{likelihood_factor}
p(\tb{y},\bm{\tau}, \bm{\gamma}|\tb{d},\bm{\theta},\tb{v},\bm{\beta})& = \prod_{n=1}^N  \Big[ p(\gamma_{n}|\theta_{n},v_{n}) \nonumber\\&p(\tau_{n}|d_{n})\prod_{l=1}^{N_r} p(y_{n}^{[l]}|\theta_{n},\beta_n)
\Big],
\end{align}
where $p(\tau_{n}|d_{n})\propto\mathcal{N}(r_n;\frac{2 d_n}{c},\sigma_\tau^2)$ and $p(\gamma_{n}|\theta_{n},v_{k,n})\propto\mathcal{N}(\gamma_n;\frac{2v_{n} \cos \theta_{n} f_c}{c},\sigma_\gamma^2)$. Recalling the model \eqref{model}, the received signal $y_{n}^{[l]}$ at the $l$th receive antenna consists of $N_t$ components, which makes the inference problem very difficult. Hence we introduce an auxiliary variable $\epsilon_{n}^{[q]}$ satisfying $\epsilon_{n}^{[q]}=e^{-j\pi q \cos\theta_{n}}$ and $y_{k,n}^{[l]}=\sum_{i=1}^{N_t} a_i({\hat{\theta}}_{n}^{0}) \epsilon_{n}^{[i-l]}+z_{y}$. Based on the auxiliary variables, $p(y_{n}^{[l]}|\theta_{n})$ is given by
\begin{align}\label{likelihood_ykn}
p(y_{n}^{[l]}|\theta_{n},\beta_n) \propto &\exp\left(\frac{|y_{n}^{[l]}-\beta_n\sqrt{p_n}\sum_{i=1}^{N_t} a_i({\hat{\theta}}_{n}^{0}) \epsilon_{n}^{[i-l]}|^2}{2\sigma_y^2}\right)\nonumber\\&\cdot\underbrace{\delta(\epsilon_{n}^{[i-l]}-e^{-j\pi (i-l) \cos\theta_{n}})}_{\kappa_{i-l}}.
\end{align}

\begin{figure}[!t]
	\centering
	\includegraphics[width=.5\textwidth]{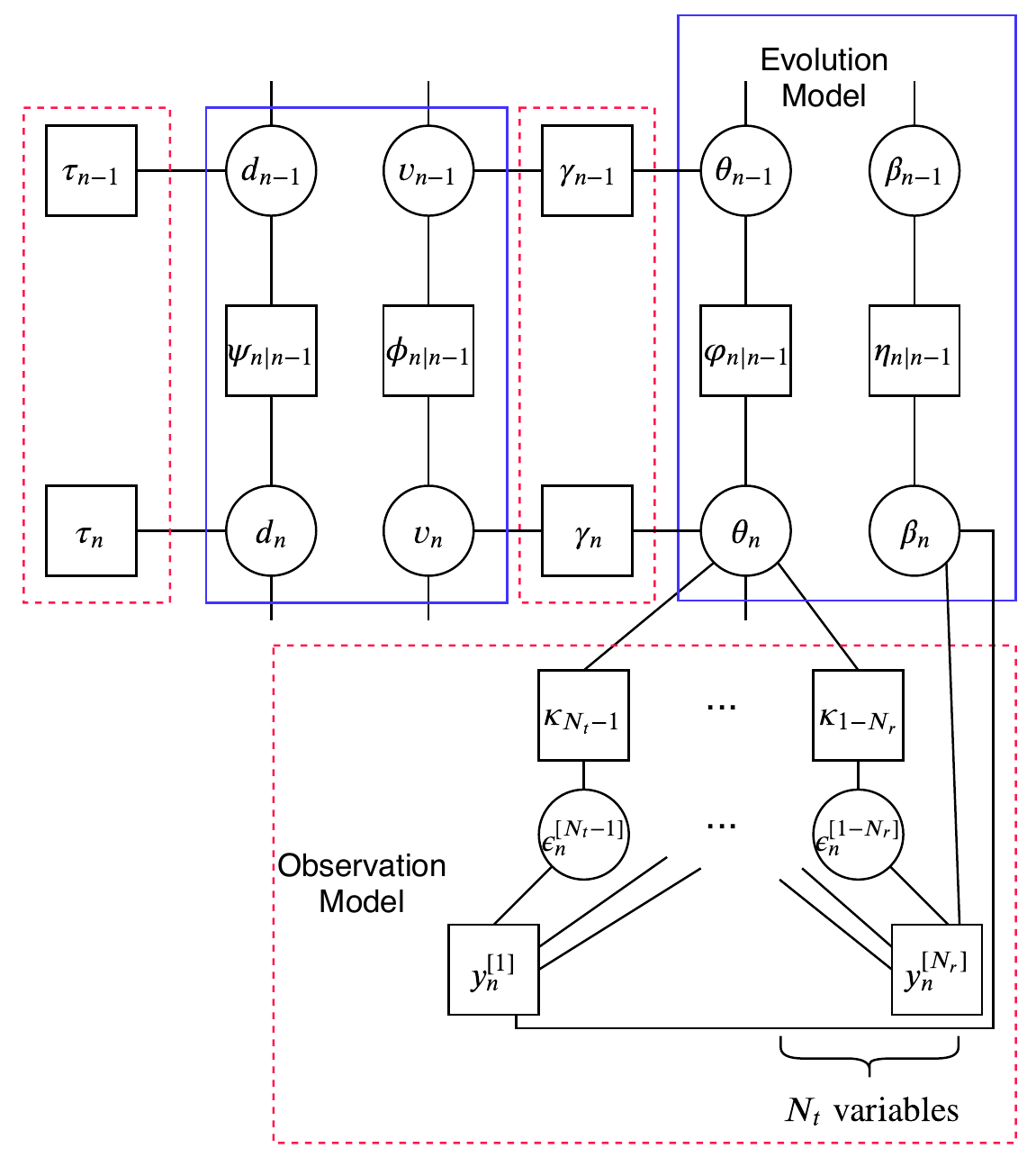}
	\caption{The factor graph representation of the considered problem.}\label{fig1}
	\centering
\end{figure}

Following \eqref{post_factor}-\eqref{likelihood_ykn}, we have the factorization of the joint \emph{a posteriori} distribution and can represent it by a factor graph, as shown in Fig. \ref{fig1}, where each square represents an factor vertex and each circle denotes a variable vertex. In Fig. \ref{fig1}, the blue solid line boxed area and red dashed line boxed area correspond to the state evolution model and observation model, respectively. The shorthand notations $\psi_{n|n-1}$, $\phi_{n|n-1}$, $\varphi_{n|n-1}$, and $\eta_{n|n-1}$ denote the state transition probabilities $p(d_{n}|d_{n-1})$, $p(v_{n}|v_{n-1})$, $p(\theta_{n}|\theta_{n-1})$, and $p(\beta_n|\beta_{n-1})$, respectively. The factor vertices $\tau_{n}$, $\gamma_{n}$, and $y_n^{[i]}$ denote the likelihood functions corresponding to the observations $\tau_{n}$, $f_{n}$ and $y_n^{[i]}$.
Having the factor graph, message passing algorithm can be implemented to efficiently compute the ``beliefs'' (approximate marginals) of unknown variables, which will be elaborated in the following section.

\section{The Proposed Message Passing Approach}
This section presents the proposed message passing algorithm for predicting and tracking the beam. There are two kinds of messages, i.e., the message from the factor vertex to the variable vertex and vice versa. For notational convenience, we use $\rmsg{\mu}_{f}(x)$ to denote the message from the factor vertex $f$ to the variable vertex $x$ and $\lmsg{\nu}_{f}(x)$ to denote the message from $x$ to factor $f$.
\subsection{Conventional Message Passing Algorithm}
The conventional message passing algorithm, also known as belief propagation (BP) has defined the following message updating rules,
\begin{align}\label{ftox_scma_ftn}
\rmsg{\mu}_{f}(x)&\propto\int 	f(\tb{x}) \prod _{x'\in \mathcal{S}_f\backslash \{x\}}\lmsg{\nu}_{ f} (x') \textrm{d} {x}',\\
\label{xtof_scma_ftn}
\lmsg{\nu}_{f}(x)&\propto\prod_{f'\in \mathcal{S}_x\backslash\{f\}} \rmsg{\mu}_{f'} (x),
\end{align}
where $\mathcal{S}_f$ denotes the set of variables in function $f(\tb{x})$ and $\mathcal{S}_x$ denotes the set of factor vertices connected to $x$. Having obtained all messages from $\forall f \in \mathcal{S}_f$ to variable vertex $x$, the belief (marginal) of variable $x$ is then given by
\begin{align}\label{belief_x}
b(x)=\prod_{f\in \mathcal{S}_x}\rmsg{\mu}_{f} (x).
\end{align}
\subsection{Vehicle State Prediction}
As discussed in Section II, the RSU and the vehicles design the receive beamformers based on the predicted angles. With the estimates of the motion parameters at time $n-1$, the RSU can predict angle $\theta_n$ via the state evolution model. Furthermore, the RSU can perform a prediction of the angle at time $n+1$ based on the information obtained at the $(n-1)$th time instant. In this way, the information of the predicted angle $\theta_{n+1}^{\rm pred}$ is contained in the DFRC signal sent to the vehicles. After receiving the angle information, the vehicles can formulate the receive beamformer $\tb{w}_{n+1}$ at the $(n+1)$th instant using the predicted angles $\theta_{n+1}^{\rm pred}$ for in formation reception at time $n+1$. If the prediction is sufficiently accurate, the transmit beams of RSU and the receive beams of vehicles are aligned, which leads to a better communication performance.

We commence our discussions with the messages in the state evolution part. Provided that the belief of $v_{n-1}$ has been obtained in Gaussian form as $b(v_{n-1})=\mathcal{N}(v_{n-1}; m_{v_{n-1}},\lambda_{{v}_{n-1}})$, the message $\rmsg{\mu}_{\phi_{n|n-1}}(v_n)$ is calculated according to \eqref{ftox_scma_ftn}, given by
\begin{align}
&\rmsg{\mu}_{\phi_{n|n-1}}(v_n)\propto \nonumber\\&\int  \exp\left(-\frac{(v_n-v_{n-1})^2}{2\sigma_v^2}\right)\exp\left(\frac{(v_{n-1}-m_{v_{n-1}})^2}{2\lambda_{{v}_{n-1}}}\right) \textrm{d} v_{n-1},\nonumber\\
&\hspace{18mm}\propto \exp\left(-\frac{(v_n-m_{v_{n-1}})^2}{2(\sigma_v^2+\lambda_{{v}_{n-1}})}\right).
\end{align}
It can be observed that the above message subjects to Gaussian distribution, which is characterized by the mean ${m}_{\phi_{n|n-1}\to v_n}=m_{v_{n-1}}$ and variance ${\lambda}_{\phi_{n|n-1}\to v_n}=\sigma_v^2+\lambda_{{v}_{n-1}}$. Therefore, we use the corresponding mean and variance to simplify the message derivations.

In a similar way, we can derive the messages $\rmsg{\mu}_{\psi_{n|n-1}}(d_n)$, $\rmsg{\mu}_{\eta_{n|n-1}}(\beta_n)$, and $\rmsg{\mu}_{\varphi_{n|n-1}}(\theta_n)$ related to the vehicle predication based on the transition probabilities \eqref{transition_2}-\eqref{transition_4}, expressing as
\begin{align}\label{eq44}
\left\{
\begin{array}{l}
{m}_{\psi_{n|n-1} \to d_n}=m_{d_{n-1}}-\hat{v}_{n-1} T\cos \hat{\theta}_{n-1},\\
{\lambda}_{\psi_{n|n-1} \to d_n}=\sigma_d^2+\lambda_{d_{n-1}},\\
{m}_{\varphi_{n|n-1}\to\theta_n}=m_{\theta_{n-1}}+\frac{\hat{v}_{n-1} T\sin \hat{\theta}_{n-1}}{\hat{d}_{n-1}},\\
{\lambda}_{\varphi_{n|n-1}\to\theta_n}=\sigma_\theta^2+\lambda_{\theta_{n-1}},\\
{m}_{\eta_{n|n-1}\to\beta_n}=\rho_{n-1}\rho_{n-1} m_{\beta_{n-1}},\\
{\lambda}_{\eta_{n|n-1}\to\beta_n}=\sigma_\beta^2+\rho_{n-1}^2 \lambda_{\beta_{n-1}}.
\end{array}\right.
\end{align}
It can be observed that the means and variances in \eqref{eq44} are updated based on the marginal mean and variance in the previous time instant and the state evolution model.
Based on the Gaussian form message $\rmsg{\mu}_{\varphi_{n|n-1}}(\theta_n)$, we have the predicted angle ${\theta}^{\rm pred,R}_{n}$ at the $n$th epoch as
\begin{align}\label{predRSU}
{\theta}^{\rm pred,R}_{n}=\arg\max_\theta \rmsg{\mu}_{\varphi_{n|n-1}}(\theta_n) = {m}_{\varphi_{n|n-1}\to\theta_n},
\end{align}
which is used for designing the beamformer at the RSU.

In addition, to realize the predictive beamforming at all vehicles, the RSU performs a further prediction of the relative angle at time $n+1$ based on the estimate $\hat{\theta}_{n|n-1}$ as well as the predicted motion parameters ${m}_{\psi_{n|n-1}\to d_n}$ and ${m}_{\phi_{n|n-1}\to v_n}$, given by
\begin{align}
{\theta}^{\rm pred}_{n+1} = \hat{\theta}_{n|n-1}+\frac{m_{v_{n-1}} T \sin \hat{\theta}_{n|n-1}}{m_{d_{n-1}}-\hat{v}_{n-1} T\cos \hat{\theta}_{n-1}}.
\end{align}
The uncertainty of the angle is given by $\lambda_{\theta_{n+1|n-1}}= 2\sigma^2_\theta+\lambda_{\theta_{n-1}}$. The predicted angle $\theta^{\rm pred}_{n+1}$ is then sent to the vehicles for receive beam steering at time $n+1$.

\subsection{Vehicle State Tracking}
At the $n$th time instant, the vehicles receive the data information as well as the predicted angles for time $n+1$. Then the vehicles can decode the information and formulate their beamformers to receive the signals at the $(n+1)$th time instant. On the other hand, the RSU receives the echoes reflected by the vehicles. Based on the observations, the RSU is able to refine the estimates of the motion parameters at time $n$. Then following the process in the above subsection, the RSU can predict the angles for the $(n+1)$th and $(n+2)$th time instants. In the following, we will discuss the messages calculations related to the observations.
\subsubsection{Messages related to $d_n$}
The message $\rmsg{\mu}_{\tau_n} (d_n)$ is identical to the likelihood function $p(\tau_n|d_n)$ since $\tau_n$ depends solely on $d_n$. After straightforward manipulations, we write $\rmsg{\mu}_{\tau_n} (d_n)$ as
\begin{align}
\rmsg{\mu}_{\tau_n} (d_n) \propto \mathcal{N}\left(d_n;\frac{c\tau_n}{2},\frac{\sigma_\tau^2\,c^2}{4}\right).
\end{align}
Then the belief of $d_n$ at time instant $n$ can be obtained as\footnote{Note that the considered system is causal, therefore the motion parameters of vehicles depend only on the past states. The messages only forward along the time direction.}
\begin{align}\label{belief_d_n}
b(d_n) &= \rmsg{\mu}_{\tau_n} (d_n) \cdot \rmsg{\mu}_{\psi_{n|n-1}} (d_n)\nonumber\\
&=\mathcal{N}\left(d_n;m_{d_n},\lambda_{d_n}\right),
\end{align}
with the mean and variance being
\begin{align}
m_{d_n} &= \lambda_{d_n}\left(\frac{2c\tau}{\sigma_\tau^2\,c^2}+\frac{m_{d_{n-1}}-\hat{v}_{n-1} T\cos \hat{\theta}_{n-1}}{\sigma_d^2+\lambda_{d_{n-1}}}\right),\\
\lambda_{d_n} &= \left(\frac{4}{\sigma_\tau^2\,c^2}+\frac{1}{\sigma_d^2+\lambda_{d_{n-1}}}\right)^{-1}.
\end{align}
Since $b(d_n)$ is a Gaussian distribution, the estimate of range $d_n$ is $\hat{d}_{n} = m_{d_n}$. The estimate $\hat{d}_{n}$ is used for modeling the state evolution function \eqref{state_evolution_theta}. Also, the obtained belief is passed to factor vertex $\psi_{n+1|n}$ for calculating $\rmsg{\mu}_{\psi_{n|n-1}}(d_n)$.

\subsubsection{Messages related to $\gamma_n$}
\begin{figure}[]
	\centering
	\includegraphics[width=.35\textwidth]{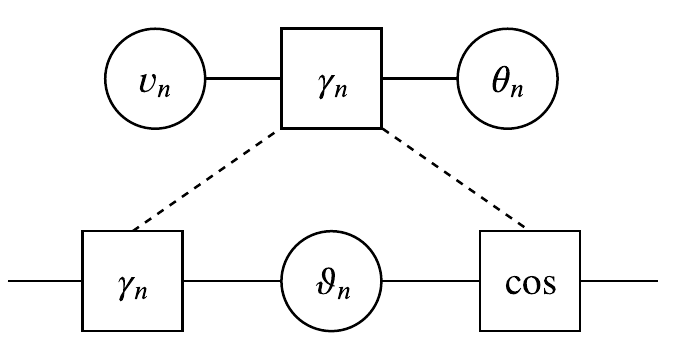}
	\caption{Reconstruction of the factor vertex $\gamma_n$.}\label{fig2}
	\centering
\end{figure}

Then, we focus on the message updating concerning the speed variable. Note that \eqref{doppler} involves a nonlinear cosine function, calculating the message $\rmsg{\mu}_{\gamma_n}(v_{n-1})$ can not provide a closed-form expression. To tackle this problem, we reconstruct the factor node $\tau_n$ by introducing a factor vertex representing the cosine function and a variable vertex denoting the cosine of an angle, as illustrated in Fig. \ref{fig2}.
Provided the message $\lmsg{\nu}_{\gamma_n}(\vartheta_{n})$ obeys $\mathcal{N}(\vartheta_{n};m_{\vartheta_{n}\to \gamma_n},\lambda_{\vartheta_{n}\to \gamma_n})$, we can now determine the message $\rmsg{\mu}_{\gamma_n}(v_{n-1})$ as
\begin{align}
\rmsg{\mu}_{\gamma_n}(v_{n}) \propto &\int \exp\left(-\frac{(\gamma_n-{ \frac{2f_c}{c} v_n\vartheta_n})^2}{2\sigma_\gamma^2}\right)\nonumber\\&\cdot \exp\left(-\frac{(\vartheta_n-m_{\vartheta_{n}\to \gamma_n})^2}{2\lambda_{\vartheta_{n}\to \gamma_n}}\right)\textrm{d}\vartheta_n \nonumber\\
\propto & \exp\left(-\frac{(\gamma_n-v_{n}m_{\vartheta_{n}\to \gamma_n})^2}{2(\sigma_\gamma^2+v_{n}^2 \lambda_{\vartheta_{n}\to \gamma_n})}\right).
\end{align}
Although we derive a closed form solution for the integration problem, it is seen that the variable $v_n$ appears in both sides of the fraction bar, which indicates that we can not write the message $\rmsg{\mu}_{\gamma_n}(v_{n})$ in a Gaussian form. Hence we resort to the MF message passing algorithm, in which the message $\rmsg{\mu}_f(x)$ is defined as
\begin{align}\label{ftox_vmp}
\rmsg{\mu}_{f}( x) &\propto\exp\left(\int \ln f(\tb{x})\prod _{x'\in \mathcal{S}_f\backslash \{x\}}\lmsg{\nu}_{ f} (x') \textrm{d} {x}'\right).
\end{align}
Following \eqref{ftox_vmp}, we can derive the Gaussian message $\rmsg{\mu}_{\gamma_n}(v_{n})$, whose mean and variance are
\begin{align}
m_{\gamma_n\to v_n} = \frac{\gamma_n m_{\vartheta_{n}\to \gamma_n}}{C_1(\lambda_{\vartheta_{n}\to \gamma_n}+m_{\vartheta_{n}\to \gamma_n}^2)},~C_1 = \frac{2 f_c}{c}\\
\lambda_{\gamma_n \to v_n} = \frac{\sigma_\gamma^2}{C_1^2(\lambda_{\vartheta_{n}\to \gamma_n}+m_{\vartheta_{n}\to \gamma_n}^2)},
\end{align}
respectively. Similarly, the mean and variance of $\rmsg{\mu}_{\gamma_n}(\vartheta_{n})$ are given by
\begin{align}
m_{\gamma_n\to \vartheta_n} = \frac{\gamma_n m_{v_{n}\to \gamma_n}}{C_1(\lambda_{v_{n}\to \gamma_n}+m_{v_{n}\to \gamma_n}^2)},\\
\lambda_{\gamma_n \to \vartheta_n} = \frac{\sigma_\gamma^2}{C_1^2(\lambda_{v_{n}\to \gamma_n}+m_{v_{n}\to \gamma_n}^2)},
\end{align}
respectively, where $m_{v_{n}\to \gamma_n}$ and $\lambda_{v_{n}\to \gamma_n}$ are identical to ${m}_{\phi_{n|n-1}\to v_n}$ and ${\lambda}_{\phi_{n|n-1}\to v_n}$, respectively.
The belief of $v_n$ can be obtained with message $\rmsg{\mu}_{\gamma_n}(v_{n})$ and $\rmsg{\mu}_{\phi_{n|n-1}}(v_n)$ and the estimate $\hat{v}_n$ is used for predicating the parameters in the $(n+1)$th epoch.

Next we move our focus to the nonlinear cosine function that maps the angle $\theta_n$ to a real number $\vartheta_n$. According to the characteristic function \cite{feuerverger1977empirical}, the expectation of $e^{j\theta}$ given $\theta\sim\mathcal{N}(\theta;0,\lambda_\theta)$ is $\mathbb{E}[e^{j\theta}] = e^{0+(j\theta)^2/2} = e^{-\lambda_\theta/2}$. It is well known that $e^{j\theta}=\cos\theta+j \sin\theta$, hence we have $\mathbb{E}[\cos\theta]= e^{-\lambda_\theta/2}$ while $\mathbb{E}[\sin\theta]= 0$. Considering the message $\lmsg{\nu}_{\cos}(\theta_n) = \mathcal{N}(\theta_n;{m}_{\theta_n\to\cos},{\lambda}_{\theta_n\to\cos})$, we can obtain parameters $m_{\vartheta_{n}\to \gamma_n}$ and $\lambda_{\vartheta_{n}\to \gamma_n}$ as
\begin{align}\label{mcostheta}
m_{\vartheta_{n}\to \gamma_n}&= \mathbb{E}[\cos \theta_n]= \mathbb{E}[\cos (\theta+{m}_{\theta_n\to\cos})] \nonumber\\&=\mathbb{E}[\cos\theta \cos {m}_{\theta_n\to\cos}-\sin\theta\sin {m}_{\theta_n\to\cos}] \nonumber\\&= e^{-{\lambda}_{\theta_n\to\cos}/2}\cos {m}_{\theta_n\to\cos},\\\label{vcostheta}
\lambda_{\vartheta_{n}\to \gamma_n} &= \mathbb{E}[\cos^2 \theta_n]-\mathbb{E}[\cos \theta_n]^2 \nonumber\\&= \frac{1}{2}\Big(\cdot e^{-2{\lambda}_{\theta_n\to\cos}}\cos 2{m}_{\theta_n\to\cos}\Big)\nonumber\\&-e^{-{\lambda}_{\theta_n\to\cos}}\cos^2 {m}_{\theta_n\to\cos}.
\end{align}
For message $\rmsg{\mu}_{\cos}(\theta_n)$, the function turns to be an inverse cosine function. To overcome this nonlinear issue, one can directly employ the particle filtering (PF) based method by representing the nonlinear distributions by particles. However, the huge computational cost violates the low-latency requirement of vehicular applications. As an alternative, we employ the second order Taylor expansion concerning the inverse cosine function as $\arccos\vartheta\approx \pi/2-\vartheta-\vartheta^3/6$. Then based on the obtained parameter $m_{\gamma_n\to \vartheta_n}$ and $\lambda_{\gamma_n \to \vartheta_n}$, we derive the Gaussian message $\rmsg{\mu}_{\gamma_n}(\theta_n)$ as
\begin{align}\label{mcostotheta}
m_{\cos \to \theta_n} &= \mathbb{E}[\arccos\vartheta_n] = \frac{\pi}{2}-\mathbb{E}[\vartheta_n]-\frac{\mathbb{E}[\vartheta_n^3]}{6}\nonumber\\
&\hspace{-10mm}=\frac{\pi}{2}-m_{\gamma_n\to \vartheta_n}-\frac{m_{\gamma_n\to \vartheta_n}^3+3m_{\gamma_n\to \vartheta_n}\lambda_{\gamma_n \to \vartheta_n}}{6},\\\label{vcostotheta}
\lambda_{\cos \to \theta_n} &= \mathbb{E}[\arccos^2\vartheta_n]- m_{\cos \to \theta_n}^2.
\end{align}
The detailed expression of $\lambda_{\cos \to \theta_n}$ is not given here, which involves the sixth-order moment of $\vartheta_n$. Please refer to the generalized Hermite polynomials for the expressions of higher order moments of $\vartheta_n$. 
%
\subsubsection{Messages related to $y_{n}^{[l]}$}
Next, we will derive the messages related to the observations $y_{n}^{[l]}$. Assuming that all messages from $\epsilon_{n}^{[q]}$ to $y_{n}^{[l]}$ are known with Gaussian distributions, it is readily to obtain the message $\rmsg{\mu}_{y_{n}^{[l]}}(\beta_n)$ using MF rules as
\begin{align}
m_{y_{n}^{[l]} \to \beta_n}&=\frac{m^{*}_{\epsilon\to y_{n}^{[l]}}y_n^{[l]}}{\sqrt{p_n} (v_{\epsilon\to y_{n}^{[l]}}+|m_{\epsilon\to y_{n}^{[l]}}|^2)},\\
\lambda_{y_{n}^{[l]} \to \beta_n}&=\frac{\sigma_y^2}{p_n (v_{\epsilon\to y_{n}^{[l]}}+|m_{\epsilon\to y_{n}^{[l]}}|^2)},
\end{align}
where $m_{\epsilon\to y_{n}^{[l]}}= \sum_{i=1}^{N_t} a_i(\hat{\theta}_{n|n-1}) m_{\epsilon_{n}^{[i-l]} \to y_{n}^{[l]}}$, \text{and}~$\lambda_{\epsilon\to y_{n}^{[l]}}= \sum_{i=1}^{N_t} |a_i(\hat{\theta}_{n|n-1})|^2 v_{\epsilon_{n}^{[i-l]} \to y_{n}^{[l]}}$.
The mean $m_{y_{n}^{[l]} \to \beta_n}$ and $\lambda_{y_{n}^{[l]} \to \beta_n}$ represents the information of the observation at the $l$th receive antenna contributed to the variable $\beta_n$. Having $\rmsg{\mu}_{y_{n}^{[l]}}(\beta_n)$ in hand, the belief of $\beta_n$ can be obtained according to \eqref{belief_x} with mean and variance
\begin{align}
m_{\beta_n} &= v_{\beta_n}\left(\frac{m_{\eta_{n|n-1}\to\beta_n}}{\lambda_{\eta_{n|n-1}\to\beta_n}}+\sum_{l=1}^{N_r}\frac{m_{y_{n}^{[l]} \to \beta_n}}{\lambda_{y_{n}^{[l]} \to \beta_n}}\right),\\
\lambda_{\beta_n} &= \left(\frac{1}{\lambda_{\eta_{n|n-1}\to\beta_n}}+\sum_{l=1}^{N_r}\frac{1}{\lambda_{y_{n}^{[l]} \to \beta_n}}\right)^{-1}.
\end{align}
Obviously, the update of $m_{\beta_n}$ and $\lambda_{\beta_n}$ depend on the state evolution information from the previous time instant as well as the information from the $N_r$ receive antennas. Consequently, the message from $\beta_n$ to a factor vertex $y_{n}^{[l]}$ is simply derived by
\begin{align}
m_{\beta_n\to y_{n}^{[l]}}&=\frac{\lambda_{\beta_n\to y_{n}^{[l]}}m_{\beta_n}-\lambda_{\beta_n}m_{y_{n}^{[l]} \to \beta_n}}{v_{\beta_n\to y_{n}^{[l]}}-\lambda_{\beta_n}},\\
\lambda_{\beta_n\to y_{n}^{[l]}}& = \frac{\lambda_{\beta_n\to y_{n}^{[l]}}\lambda_{\beta_n}}{v_{\beta_n\to y_{n}^{[l]}}-\lambda_{\beta_n}}.
\end{align}
Provided the Gaussian message $\rmsg{\mu}_{\beta_n}( y_{n}^{[l]})$, the message $\rmsg{\mu}_{y_n^{[l]}}(\epsilon_n^{[q]}),~q\in[1-l,N_t-l]$ is computed as,
\begin{align}\label{m_yn_epsilon}
&\hspace{-4mm}m_{y_n^{[l]}\to \epsilon_n^{[q]}}= \frac{y_n^{[l]}m^{*}_{\beta_n\to y_{n}^{[l]}}-\!\!\!\sum\limits_{i\neq q+l} \!\!\!a_{i}(\hat{\theta}_{n|n-1})m_{\epsilon_{n}^{[i-l]} \to y_{n}^{[l]}}}{a_{q+l}(\hat{\theta}_{n|n-1})\cdot(|m_{\beta_n\to y_{n}^{[l]}}|^2+\lambda_{\beta_n\to y_{n}^{[l]}})},\\\label{v_yn_epsilon}
&\hspace{-4mm}\lambda_{y_n^{[l]}\to \epsilon_n^{[q]}}=\frac{\sigma^2_y}{|a_{q+l}(\hat{\theta}_{n|n-1})|^2\cdot(|m_{\beta_n\to y_{n}^{[l]}}|^2+\lambda_{\beta_n\to y_{n}^{[l]}})}.
\end{align}
Using \eqref{m_yn_epsilon} and \eqref{m_yn_epsilon}, the information obtained from the observation $y_n^{[l]}$ is fed to the auxiliary variable and then to the angular parameter.

\subsubsection{Messages related to $\kappa_q$}
\begin{figure}[]
	\centering
	\includegraphics[width=.5\textwidth]{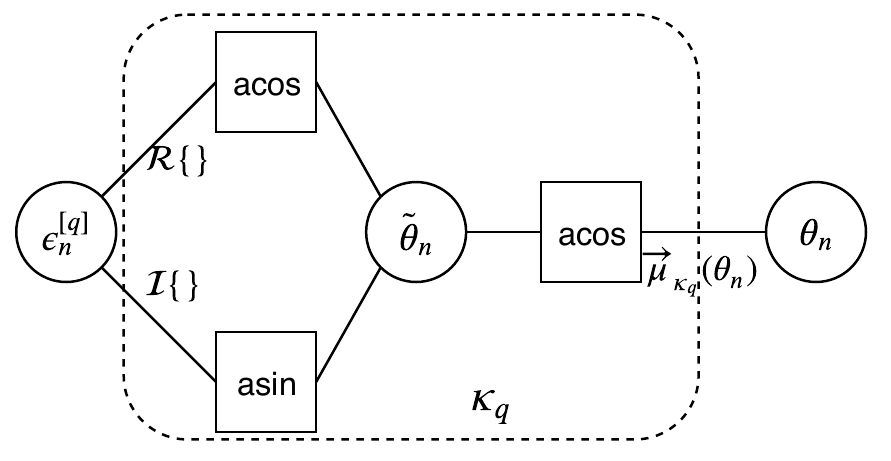}
	\caption{Reconstruction of the factor vertex $\kappa_q$.}\label{fig3}
	\centering
\end{figure}

Finally, we aim for computing the messages related to the function $\kappa_q$, which involves the nonlinear function $e^{-jq\cos\theta_n}$. Note that this part of factor graph has cycles, we have to implement the message passing algorithms for a few iterations\cite{ihler2005loopy}. As above, the mean and variance of $\tilde{\theta}_n=\cos\theta_n$ given the Gaussian form input message $\lmsg{\nu}_{\kappa_q}(\theta_n)$ can be calculated similar to \eqref{mcostheta} and \eqref{vcostheta}, denoting by ${m}_{\tilde{\theta}_n\to \kappa_q}$ and ${\lambda}_{\tilde{\theta}_n\to \kappa_q}$. Employing the transformations of trigonometric functions and after some manipulations, we have the mean and variance for message $\rmsg{\mu}_{\kappa_q}(\epsilon_n^{[q]})$, formulating as
\begin{align}\label{m_kappa}
m_{\kappa_q \to \epsilon_n^{[q]}} &=e^{-{q^2 {\lambda}_{\tilde{\theta}_n\to \kappa_q}}/{2}}e^{-j q {m}_{\tilde{\theta}_n\to \kappa_q}},\\\label{v_kappa}
\lambda_{\kappa_q \to \epsilon_n^{[q]}} &= 1-e^{-q^2 {\lambda}_{\tilde{\theta}_n\to \kappa_q}}.
\end{align}
It is then straightforward to determine the message $\lmsg{\nu}_{y_{n}^{[l]}}(\epsilon_n^{[q]})$ and $\lmsg{\nu}_{\kappa_{q}}(\epsilon_n^{[q]})$ as\footnote{It may happen parial messages $\rmsg{\mu}_{y_{n}^{[l']}}(\epsilon_n^{[q]})$ do not exist. In this circumstance, we cam simply set $\rmsg{\mu}_{y_{n}^{[l']}}(\epsilon_n^{[q]})=1$.}
\begin{align}\label{epstoy}
\lmsg{\nu}_{y_{n}^{[l]}}(\epsilon_n^{[q]}) &= \rmsg{\mu}_{\kappa_q}(\epsilon_n^{[q]}) \prod_{l'\neq 1,l'=1}^{N_t} \rmsg{\mu}_{y_{n}^{[l']}}(\epsilon_n^{[q]}),\\\label{epstokappa}
\lmsg{\nu}_{\kappa_{q}}(\epsilon_n^{[q]}) &= \prod_{l=1}^{N_t} \rmsg{\mu}_{y_{n}^{[l]}}(\epsilon_n^{[q]}).
\end{align}
The above two messages are still Gaussian since all messages on the right-hand-side of \eqref{epstoy} and \eqref{epstokappa} are represented in Gaussian closed-form. Computing the backward message $\rmsg{\mu}_{\kappa_q}(\theta_n)$ involves the inverse function of $e^{-jq\tilde{\theta}_n}=e^{-jq \cos\theta_n}$, which is not applicable to deliver Gaussian distributed variable $\theta_n$. To this end, we reconstruct the function node $\kappa_n$, as shown in Fig. \ref{fig3}, on which the process for calculating $\rmsg{\mu}_{\kappa_q}(\theta_n)$ is given in the following.
The real and imaginary parts of the mean as well as the variance of $\lmsg{\nu}_{\kappa_{q}}(\epsilon_n^{[q]})$ are passed to the inverse cosine (``acos'') and inverse sine (``asin'') functions, respectively. Then the mean and variance of message from the acos factor vertex to $\tilde{\theta}_n$ is calculated based on Taylor expansion similar to \eqref{mcostotheta} and \eqref{vcostotheta}. The message related to asin function is calculated according to the second order Taylor expansion of inverse trigonometric function, given by $\arcsin{\epsilon} \approx \epsilon-\epsilon^3/6$. Then the mean of the message $\rmsg{\mu}_{\text{asin}}(\tilde{\theta}_n)$ is expressed as
\begin{align}
m_{\text{asin}\to\tilde{\theta}_n} = \mathcal{I}\{m_{\epsilon_n^{[q]}\to \kappa_q}\}\left(1+\frac{\lambda_{\epsilon_n^{[q]}\to \kappa_q}}{2} + \frac{\mathcal{I}\{m_{\epsilon_n^{[q]}\to \kappa_q}\}^2}{6}\right),
\end{align}
and the expression of variance $\lambda_{\text{asin}\to\tilde{\theta}_n}$ having higher order moments is not given here. Consequently, the message $\lmsg{\nu}_{\text{acos}}(\tilde{\theta}_n)$ is simply calculated by
\begin{align}
\lmsg{\nu}_{\text{acos}}&(\tilde{\theta}_n)= \rmsg{\mu}_{\text{asin}}(\tilde{\theta}_n) \rmsg{\mu}_{\text{acos}}(\tilde{\theta}_n)\nonumber\\
&\propto\mathcal{N}\Bigg(\tilde{\theta}_n;\frac{m_{\text{acos}\to\tilde{\theta}_n}\lambda_{\text{asin}\to\tilde{\theta}_n}+m_{\text{asin}\to\tilde{\theta}_n}\lambda_{\text{acos}\to\tilde{\theta}_n} }{\lambda_{\text{asin}\to\tilde{\theta}_n} +\lambda_{\text{acos}\to\tilde{\theta}_n} }\nonumber\\&\hspace{34mm},\frac{\lambda_{\text{asin}\to\tilde{\theta}_n}\lambda_{\text{acos}\to\tilde{\theta}_n} }{\lambda_{\text{asin}\to\tilde{\theta}_n} +\lambda_{\text{acos}\to\tilde{\theta}_n} }\Bigg).
\end{align}
In the end, the message $\rmsg{\mu}_{\kappa_q}(\theta_n)$ is calculated in Gaussian form by implementing the Taylor expansion of inverse cosine function again, which has a similar form to that of \eqref{mcostotheta} and \eqref{vcostotheta}. Having determined all messages form $\kappa_q$ to $\theta_n$, we arrive at the belief of $\theta_n$,
\begin{align}
b(\theta_n) = \rmsg{\mu}_{\gamma_n}(\theta_n)\rmsg{\mu}_{\varphi_{n|n-1}}(\theta_n)\prod_q \rmsg{\mu}_{\kappa_q}(\theta_n),
\end{align}
which is used for obtaining the message $\rmsg{\mu}_{\varphi_{n+1|n}}(\theta_{n+1})$ and for beam prediction at $(n+1)$th time instant. The estimate of $\theta_n$, given by $\hat{\theta}_n=\arg\max_{\theta_n}b(\theta_n)$, is used for constructing the state transition model.

In the above, we have solved the beam prediction and beam tracking problems based on the factor graph framework. With the help of the determined angular parameters, the RSU and the vehicle can maintain a reliable link for data transmission.

\subsection{Beam Misalignment Analysis}
Recalling that a vehicle can receive the predicted angle $\theta^{\rm pred}_{n+1}$ at the $(n+1)$th time instant from the RSU at time instant $n$, therefore can adjust the direction of its receive beamformer based on the prediction. As we have the mean value ${\theta}^{\rm pred}_{n+1}$ as well as the uncertainty $\lambda_{{\theta}_{n+1|n-1}}$, the beam-steering direction of the vehicle at the $(n+1)$th time instant follows the Gaussian distribution $\mathcal{N}(\theta^{\rm V}_{n+1};\hat{\theta}^{\rm pred}_{n+1},\lambda_{{\theta}_{n+1|n-1}})$. On the other hand, the beam prediction at the RSU is performed after the beam tracking process at time $n$, as in \eqref{predRSU}.

In mmWave systems, the beamwidth is typically very narrow that beam misalignment may happen when the predicted angles of the RSU and at the vehicle differ from the actual angle.
To analyze the probability of beam misalignment, we use $\Delta=2\delta$ to denote the beamwidth and beams are aligned only if the vehicle's beam direction $\theta^{\rm V}_{n+1}$ satisfying $|\theta^{\rm V}_{n+1}-{\theta}_{n+1}|\leq\delta$ and the RSU's beam direction $\theta^{\rm R}_{n+1}$ satisfying $|\theta^{\rm R}_{n+1}-{\theta}_{n+1}|\leq\delta$, where $\theta_{n+1}$ denotes the actual angle of the vehicle relative to the RSU's array and $\delta = \pi/N_{\rm antenna}$ \cite{zhang2019multiple}. This is equivalent to calculating the probability of
\begin{align}\label{misalign}
p_{\textrm{mis}} &= 1-p_{\textrm{alignV}}p_{\textrm{alignR}}, ~{\rm where}\\
p_{\textrm{alignV}} &= \textrm{Pr}\{\theta^{\rm V}_{n+1}\leq{\theta}_{n+1}+\delta\}-\textrm{Pr}\{\theta^{\rm V}_{n+1}\leq{\theta}_{n+1}-\delta\},\nonumber\\
p_{\textrm{alignR}} &= \textrm{Pr}\{\theta^{\rm R}_{n+1}\leq{\theta}_{n+1}+\delta\}-\textrm{Pr}\{\theta^{\rm R}_{n+1}\leq{\theta}_{n+1}-\delta\}.\nonumber
\end{align}
Given the distribution of $\theta^{\rm V}_{n+1}$, we have the cumulative distribution function (CDF) as
\begin{align}
\Phi_1(x) &= C_2\int_{-\infty}^{x} \exp\left(-\frac{(\theta^{\rm V}_{n+1}-{\theta}^{\rm pred}_{n+1})^2}{2\lambda_{{\theta}_{n+1|n-1}}}\right)\textrm{d} \theta^{\rm V}_{n+1}\nonumber\\
&= \frac{1}{2}\left[1+\textrm{erf}\left(\frac{x-{\theta}^{\rm pred}_{n+1}}{\lambda_{{\theta}_{n+1|n-1}}\sqrt{2}}\right)\right].
\end{align}
Similarly, we have the CDF $\Phi_2(x)$ for Gaussian distribution $\rmsg{\mu}_{\varphi_{n+1|n}}(\theta_{n+1})$ as
\begin{align}
\Phi_2(x)= \frac{1}{2}\left[1+\textrm{erf}\left(\frac{x-m_{\varphi_{n+1|n\to \theta_{n+1}}}}{\lambda_{\varphi_{n+1|n\to \theta_{n+1}}}\sqrt{2}}\right)\right].
\end{align}
Consequently, the misalignment probability $p_{\textrm{mis}}$ is calculated as
\begin{align}
p_{\textrm{mis}} = 1-(\Phi_1({\theta}_{n+1}+\delta)-\Phi_1({\theta}_{n+1}-\delta))\nonumber\\&\hspace{5mm}\cdot(\Phi_2({\theta}_{n+1}+\delta)-\Phi_2({\theta}_{n+1}-\delta)).
\end{align}
{Observed from the above analysis, the misalignment probability depends on the width of the beam, which motivates us to optimize the beamwidth in our future work to minimize the probability $p_{\textrm{mis}}$.}

\subsection{Computational Complexity and Signaling Overhead}
Observe from Sec. IV-B and IV-C that the complexity of the proposed algorithm is dominated by the calculations of corresponding messages. It is worth to see that all messages can be written in a closed-form by employing appropriate approximations and message updating rules. Hence, the message calculations only involve trivial addition and multiplication operations, leading to a very low level of complexity. To elaborate, the order of complexity of the proposed algorithm can be given by $\mathcal{O}(4K)$. In contrast, the PF-based algorithm adopts $R$ particles to represent the nonlinear functions, which has the order of complexity $\mathcal{O}(4KR)$. In general, $R$ should be sufficiently large to achieve a reasonably good estimation performance, and the complexity is significantly increased. As a commonly used scheme for beam tracking, the extended Kalman filtering (EFK) requires the matrix inversion, having a complexity order of cube of the matrix dimension, i.e., $\mathcal{O}((4K)^3)$. Through this complexity comparison, we show the superiority of the proposed algorithm in the large scale vehicular networks. Concerning the signaling overhead, the feedback-based approach usually utilizes one or two pilot symbols for beam tracking. For the proposed DFRC-based beam tracking scheme, since the whole downlink block is used for transmitting data information, the signaling overhead much lower than the feedback-based scheme.

\section{Simulation Results}
\begin{figure}[!t]
    \centering
    \includegraphics[width=0.9\columnwidth]{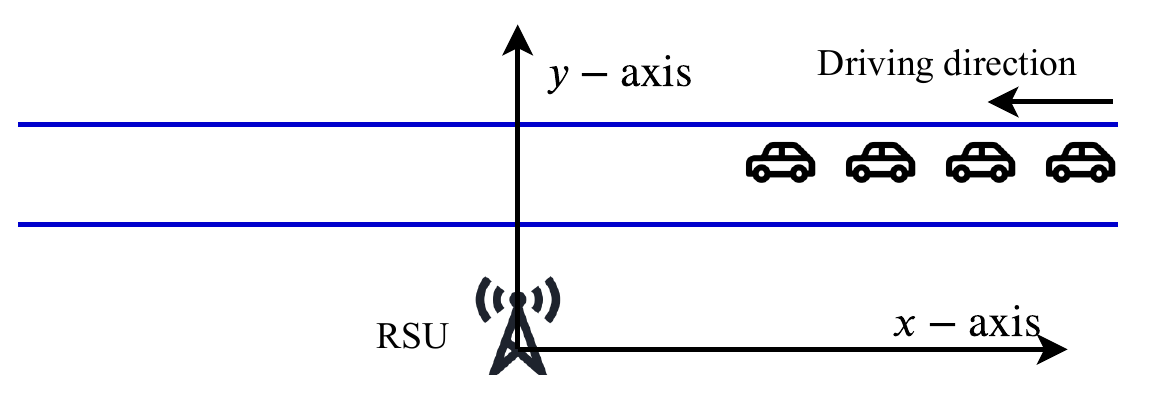}
    \caption{{The considered vehicular network for simulations.}}
    \label{simu_model}
\end{figure}
Let us consider a network with 4 vehicles moving on the road, as shown in Fig. \ref{simu_model}. Without loss of generality, the coordinate of the RSU is set as $[0,0]^{\rm T}$ and the initial positions of vehicles are $[100, 20]^{\rm T}$, $[90, 20]^{\rm T}$, $[80, 20]^{\rm T}$, and $[70, 20]^{\rm T}$, respectively. The RCS $\xi$ is set to $10+10j$, which is used for calculating the reflection coefficient $\beta_{k,0}$ via \eqref{beta_model}. The speeds of four vehicles at time instant $0$ are randomly generated from the uniform distribution $[5,20]$ m/s. The RSU and the vehicles are operating at a carrier frequency of $f_c=30$ GHz. The time slot duration is $T=0.02~s$ and the signal propagation speed is approximated as $c=3\times10^{8}$ m/s. For brevity, we set both the radar noise variance $\sigma_y^2$ and the noise PSD for communication $N_0$ to 1. For the observed delays and Dopplers at RSU, we use the standard deviations of $\sigma_\tau = 0.67~\mu$s and $\sigma_\gamma = 2$ kHz for all vehicles at different time slots. The state transition noises are set with standard deviations of $\sigma_d = 0.2$ m, $\sigma_v = 0.5$ m/s, $\sigma_\beta = 1$, and $\sigma_\theta = 0.02^{\circ}$. For the log-distance pathloss model, we assume a unit channel gain at the reference distance $d_0 = 1$ m. The maximum number of iterations for message passing algorithm is set to $10$. All results are averaged from 1000 independent Monte Carlo simulations unless otherwise specified.

\begin{figure}[!t]
    \centering
    \includegraphics[width=0.9\columnwidth]{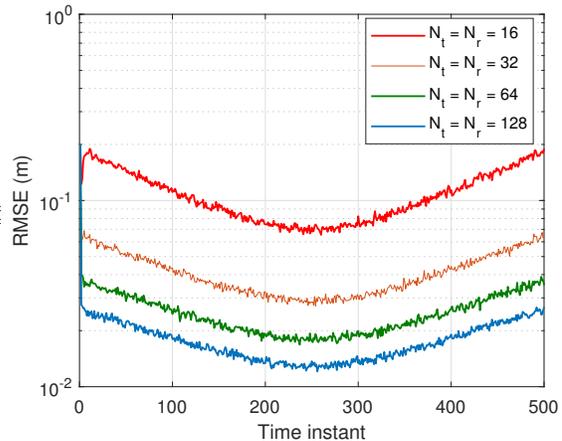}
    \caption{RMSE of the range estimation versus the time instant.}
    \label{RMSE_time}
\end{figure}
\begin{figure}[!t]
    \centering
    \includegraphics[width=0.9\columnwidth]{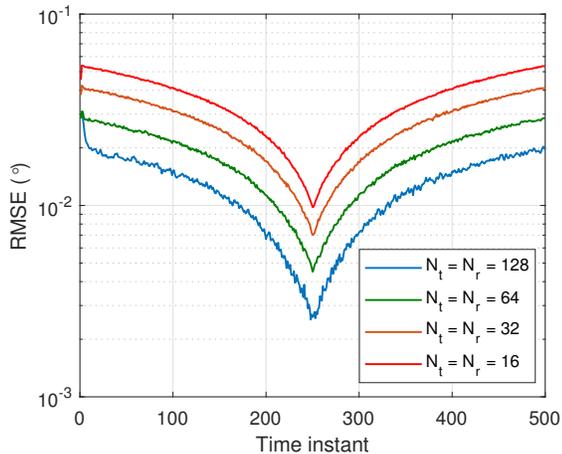}
    \caption{RMSE of the angle estimation versus the time instant.}
    \label{RMSE_angle_time}
\end{figure}

We first consider the tracking performance of the motion parameters of vehicles relying on the proposed approach. In Fig. \ref{RMSE_time}, we show the tracking result of the range parameter $d_n$ in terms of root mean squared error (RMSE). The RMSE is averaged from the results obtained for all 4 vehicles. Four deployments with different number of transmit and receive antennas are illustrated. For all cases, it is interesting to see the curves first decreases and then increases with respect to the time index. This is because when the vehicles move towards the RSU, the reflection coefficient $\beta_n$ becomes larger, leading to a higher SNR gain. On the contrary, the SNR deceases when the vehicles are moving away. Moreover, employing a larger scale antenna array will introduce a higher array gain as well as more observations, therefore resulting in a better tracking performance. The averaged RMSE of angle estimation parameterized by different number of antennas is depicted in Fig. \ref{RMSE_angle_time}. Obviously, the RMSE curves for angle tracking have a similar trend as that in Fig. \ref{RMSE_time}. It is also observed that the estimation of the angle is very accurate, at the error level of $10^{-2}$ rad.

Next we compare the performance of the proposed approach and the classic feedback-based algorithm. For fair comparison purpose, we still employ the factor graph and message passing algorithm for the feedback-based scheme\footnote{Existing feedback-based methods \cite{shaham2019fast} rely on the EKF, which in general has a worse performance than message passing algorithm due to using first-order Taylor series expansion.}. Note that in the feedback-based scheme, the pilots are contained in the downlink communication signal. Therefore, the reflection parameter in the observation model of \eqref{model} is replaced by channel gain $\alpha_{n}$, which is determined by estimating the range parameter $d_n$. In contrast to the DFRC signal that the whole block can be used as the pilots, the feedback scheme can employ only 1 or 2 pilots, leading to a much smaller SNR gain after matched filtering. For simplicity, we equivalently multiply the noise variance $\sigma_y^2$ by a constant for the feedback-based scheme.
We compare the CDF of the estimation error of $v_n$ based on the proposed approach and the feedback scheme at the last time instant with $M=N_t=N_r=64$ antennas. The high-complexity PF-based message passing algorithm and the EKF method in \cite{liu2020radar} are utilized as a benchmark. It is observed that with significantly reduced complexity, the proposed parametric message passing method can attain the performance of the PF-based one, verifying the effectiveness of applying Taylor series expansion and MF message passing. Moreover, the proposed approach significantly outperforms the feedback-based scheme with 1 and 2 pilots due to the higher SNR gain.
\begin{figure}[!t]
    \centering
    \includegraphics[width=0.9\columnwidth]{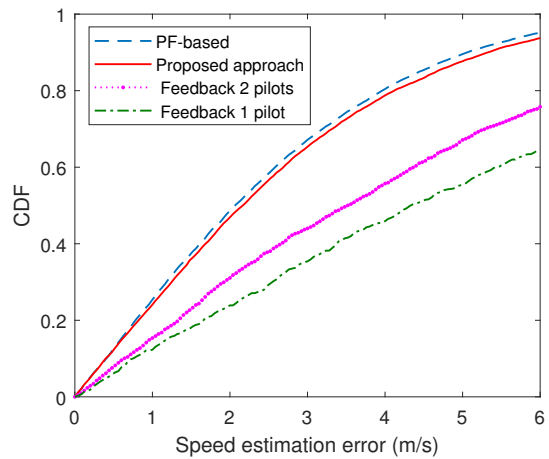}
    \caption{CDF of the speed estimation error.}\label{CDF_speed}
\end{figure}

\begin{figure}[!t]
    \centering
    \includegraphics[width=0.9\columnwidth]{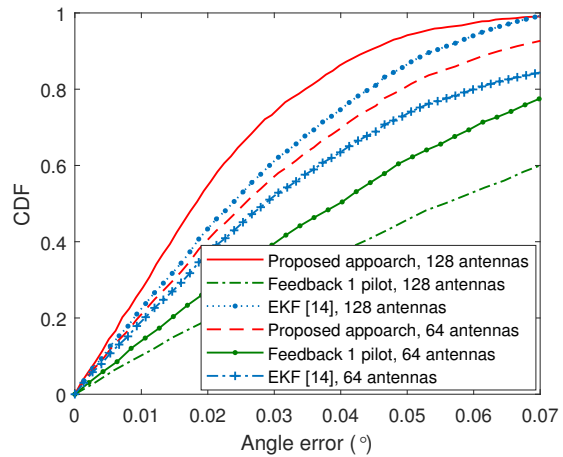}
    \caption{CDF of the angle estimation error.}\label{CDF_angle}
\end{figure}
\begin{figure}[!t]
    \centering
    \includegraphics[width=0.9\columnwidth]{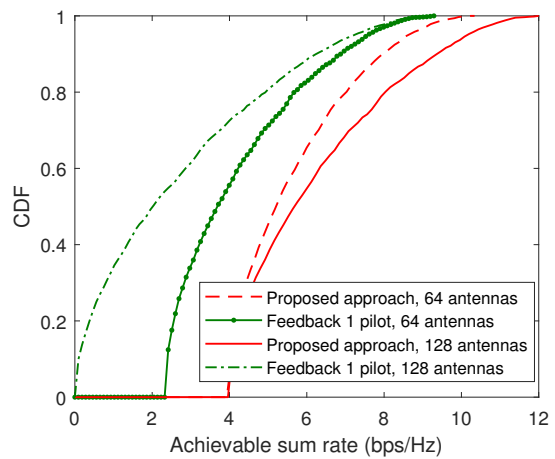}
    \caption{CDF of the communication achievable rate.}\label{CDF_rate}
\end{figure}

{Fig. \ref{CDF_angle} evaluates the angle estimation result using the proposed approach, the feedback scheme, and the EKF method in \cite{liu2020radar}. We illustrate the CDF versus the angle estimation error at the last time instant for 1000 trails. Two cases with 64 and 128 antennas are illustrated. We see that for both cases, the feedback scheme suffers from a remarkable performance loss due to limited matched-filtering gain. In contrast to the proposed approach, increasing the number of antennas for the feedback scheme will lead to performance degradation. This can be explained by the fact that a higher number of antennas provides a narrower beam, in which case using only 1 pilot is not sufficient to track the variation of the angular parameter. Moreover, the proposed algorithm outperforms the EKF method since EKF employs only the first-order Taylor expansion and neglect the higher-order information.} Since the estimated angles are used for beamforming design, the tracking error of angles will result in the misalignment of the beams. As a consequence, the received SNR is reduced, leading to a lower achievable rate. In Fig. \ref{CDF_rate}, we illustrate the CDF of the communication achievable rate of all time instants based on the proposed and the feedback-based methods at a SNR of 10 dB. For the proposed scheme, the achievable rate $R$ at different time instants are higher than 4 bps/Hz. While the achievable rate for the feedback-based scheme is much lower. This validates our discussions above that the large angle estimation error in the feedback-based approach degrades the achievable rate. Furthermore, the rate degradation becomes more significant for the feedback-based scheme in the case with 128 antennas, where the angle variation cannot be accurately tracked due to the narrow beamwidth. Figs. \ref{CDF_angle} and \ref{CDF_rate} show the superiority of employing DFRC signaling for reliable communication in vehicular networks.

\begin{figure}[!t]
    \centering
    \includegraphics[width=0.9\columnwidth]{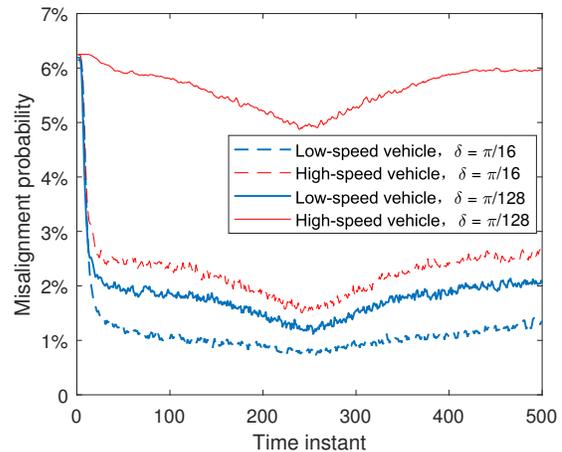}
    \caption{Beam misalignment probabilities of vehicles versus the time instant.}\label{mismatch_pro}
\end{figure}

Finally, we consider the misalignment probability with different widths of the beam. According to \eqref{misalign}, it is expected a wider beam will result in a lower misalignment probability. In Fig. \ref{mismatch_pro}, we depict the beam misalignment probabilities of two kinds of vehicles, i.e. high-speed vehicle having a speed in the interval [18, 20] m/s and low-speed vehicle having a speed in the interval [5,7] m/s, versus the time instant. The beamwidth of $\delta=\pi/16$ is achieved by activating only 16 antennas in the 128-antenna array. First we can see the misalignment probability for low-speed vehicle is lower than that for the high-speed vehicle. This is due to that the faster vehicle causes more violent variation of the relative angle $\theta_n$. In addition, we observe that the high-speed vehicle is more sensitive to the beamwidth than the low-speed one. The misalignment probability experiences a significant rise when we choose $\delta=\pi/128$. This motivates us to further optimize the beamwidth for vehicles with different speeds in practical scenarios.

\section{Conclusions}
In this paper, we proposed a novel DFRC based beamforming scheme for vehicular networks, providing zero signaling overhead for beam tracking. We commence from a Bayesian perspective and construct the joint \emph{a posteriori} distribution based on the echo signals received at the RSU and the state transition models of the vehicles. The joint distribution was further fully factorized and represented by a factor graph, then the message passing algorithm is utilized to estimate the unknown variables. A second order Taylor expansion was employed to approximate the nonlinear inverse trigonometric functions. Consequently, the messages on factor graph were determined in closed-form, providing a low complexity solution for the considered beam tracking problem. A two-step prediction of angles are further processed and sent to the vehicles via DFRC signals, which can reduce the latency for beam alignment. Simulation results demonstrate the effectiveness and superiority of the proposed approach compared to the conventional feedback-based scheme.

\bibliographystyle{IEEEtran}
\bibliography{v2x}

\begin{thebibliography}{10}
\providecommand{\url}[1]{#1}
\csname url@samestyle\endcsname
\providecommand{\newblock}{\relax}
\providecommand{\bibinfo}[2]{#2}
\providecommand{\BIBentrySTDinterwordspacing}{\spaceskip=0pt\relax}
\providecommand{\BIBentryALTinterwordstretchfactor}{4}
\providecommand{\BIBentryALTinterwordspacing}{\spaceskip=\fontdimen2\font plus
\BIBentryALTinterwordstretchfactor\fontdimen3\font minus
  \fontdimen4\font\relax}
\providecommand{\BIBforeignlanguage}[2]{{%
\expandafter\ifx\csname l@#1\endcsname\relax
\typeout{** WARNING: IEEEtran.bst: No hyphenation pattern has been}%
\typeout{** loaded for the language `#1'. Using the pattern for}%
\typeout{** the default language instead.}%
\else
\language=\csname l@#1\endcsname
\fi
#2}}
\providecommand{\BIBdecl}{\relax}
\BIBdecl

\bibitem{siegel2017survey}
J.~E. Siegel, D.~C. Erb, and S.~E. Sarma, ``A survey of the connected vehicle
  landscape—architectures, enabling technologies, applications, and
  development areas,'' \emph{IEEE Trans. Intell. Transport. Syst.}, vol.~19,
  no.~8, pp. 2391--2406, Aug. 2017.

\bibitem{lu2014connected}
N.~Lu, N.~Cheng, N.~Zhang, X.~Shen, and J.~W. Mark, ``Connected vehicles:
  Solutions and challenges,'' \emph{IEEE Internet. Things J.}, vol.~1, no.~4,
  pp. 289--299, Sep. 2014.

\bibitem{chen2017vehicle}
S.~Chen, J.~Hu, Y.~Shi, Y.~Peng, J.~Fang, R.~Zhao, and L.~Zhao,
  ``Vehicle-to-everything {(V2X)} services supported by {LTE-based systems and
  5G},'' \emph{IEEE Commun. Stand. Mag.}, vol.~1, no.~2, pp. 70--76, Jan. 2017.

\bibitem{wang2017overview}
X.~Wang, S.~Mao, and M.~X. Gong, ``An overview of {3GPP} cellular
  vehicle-to-everything standards,'' \emph{GetMobile: Mobile Comput. Commun.},
  vol.~21, no.~3, pp. 19--25, Mar. 2017.

\bibitem{skolnik2001radar}
M.~I. Skolnik, \emph{RADAR systems}.\hskip 1em plus 0.5em minus 0.4em\relax
  McGraw-Hill, NY, 2001.

\bibitem{dickmann2016automotive}
J.~Dickmann, J.~Klappstein, M.~Hahn, N.~Appenrodt, H.-L. Bloecher, K.~Werber,
  and A.~Sailer, ``Automotive radar the key technology for autonomous driving:
  From detection and ranging to environmental understanding,'' in \emph{Proc.
  IEEE Radar Conf.}\hskip 1em plus 0.5em minus 0.4em\relax IEEE, 2016, pp.
  1--6.

\bibitem{liu2020joint}
F.~Liu, C.~Masouros, A.~Petropulu, H.~Griffiths, and L.~Hanzo, ``Joint radar
  and communication design: Applications, state-of-the-art, and the road
  ahead,'' \emph{IEEE Trans. Commun.}, In press, 2020.

\bibitem{sen2010adaptive}
S.~Sen and A.~Nehorai, ``Adaptive {OFDM} radar for target detection in
  multipath scenarios,'' \emph{IEEE Trans. Signal Process.}, vol.~59, no.~1,
  pp. 78--90, Jan. 2010.

\bibitem{shi2017power}
C.~Shi, F.~Wang, M.~Sellathurai, J.~Zhou, and S.~Salous, ``Power
  minimization-based robust {OFDM} radar waveform design for radar and
  communication systems in coexistence,'' \emph{IEEE Trans. Signal Process.},
  vol.~66, no.~5, pp. 1316--1330, Mar. 2017.

\bibitem{sturm2011waveform}
C.~Sturm and W.~Wiesbeck, ``Waveform design and signal processing aspects for
  fusion of wireless communications and radar sensing,'' \emph{Proc. the IEEE},
  vol.~99, no.~7, pp. 1236--1259, Jul. 2011.

\bibitem{chiriyath2017radar}
A.~R. Chiriyath, B.~Paul, and D.~W. Bliss, ``Radar-communications convergence:
  Coexistence, cooperation, and co-design,'' \emph{IEEE Trans. Cog. Commun.
  Netw.}, vol.~3, no.~1, pp. 1--12, Jan. 2017.

\bibitem{kumari2015investigating}
P.~Kumari, N.~Gonzalez-Prelcic, and R.~W. Heath, ``Investigating the {IEEE}
  802.11 ad standard for millimeter wave automotive radar,'' in \emph{Proc.
  IEEE Veh Technol. Conf.}\hskip 1em plus 0.5em minus 0.4em\relax IEEE, 2015,
  pp. 1--5.

\bibitem{liu2018mu}
F.~Liu, C.~Masouros, A.~Li, H.~Sun, and L.~Hanzo, ``{MU-MIMO} communications
  with {MIMO} radar: From co-existence to joint transmission,'' \emph{IEEE
  Trans. Wireless Commun.}, vol.~17, no.~4, pp. 2755--2770, Apr. 2018.

\bibitem{liu2018toward}
F.~Liu, L.~Zhou, C.~Masouros, A.~Li, W.~Luo, and A.~Petropulu, ``Toward
  dual-functional radar-communication systems: Optimal waveform design,''
  \emph{IEEE Trans. Signal Process.}, vol.~66, no.~16, pp. 4264--4279, Aug.
  2018.

\bibitem{sun2014mimo}
S.~Sun, T.~S. Rappaport, R.~W. Heath, A.~Nix, and S.~Rangan, ``{MIMO} for
  millimeter-wave wireless communications: Beamforming, spatial multiplexing,
  or both?'' \emph{IEEE Commun. Mag.}, vol.~52, no.~12, pp. 110--121, Dec.
  2014.

\bibitem{venugopal2016device}
K.~Venugopal, M.~C. Valenti, and R.~W. Heath, ``Device-to-device millimeter
  wave communications: Interference, coverage, rate, and finite topologies,''
  \emph{IEEE Trans. Wireless Commun.}, vol.~15, no.~9, pp. 6175--6188, Sep.
  2016.

\bibitem{kumari2019adaptive}
P.~Kumari, S.~A. Vorobyov, and R.~W. Heath, ``Adaptive virtual waveform design
  for millimeter-wave joint communication-radar,'' \emph{IEEE Trans. Signal
  Process.}, vol.~68, pp. 715--730, Jan. 2020.

\bibitem{haghighatshoar2016beam}
S.~Haghighatshoar and G.~Caire, ``The beam alignment problem in mmwave wireless
  networks,'' in \emph{Proc. Asilomar Conf.}\hskip 1em plus 0.5em minus
  0.4em\relax IEEE, Jul. 2016, pp. 741--745.

\bibitem{va2016beam}
V.~Va, H.~Vikalo, and R.~W. Heath, ``Beam tracking for mobile millimeter wave
  communication systems,'' in \emph{Proc. IEEE Global Conf. Signal. Inf.
  Process.}\hskip 1em plus 0.5em minus 0.4em\relax IEEE, 2016, pp. 743--747.

\bibitem{zhang2019codebook}
D.~Zhang, A.~Li, M.~Shirvanimoghaddam, P.~Cheng, Y.~Li, and B.~Vucetic,
  ``Codebook-based training beam sequence design for millimeter-wave tracking
  systems,'' \emph{IEEE Trans. Wireless Commun.}, vol.~18, no.~11, pp.
  5333--5349, Nov. 2019.

\bibitem{shaham2019fast}
S.~Shaham, M.~Ding, M.~Kokshoorn, Z.~Lin, S.~Dang, and R.~Abbas, ``Fast channel
  estimation and beam tracking for millimeter wave vehicular communications,''
  \emph{IEEE Access}, vol.~7, pp. 141\,104--141\,118, 2019.

\bibitem{gonzalez2016radar}
N.~Gonz{\'a}lez-Prelcic, R.~M{\'e}ndez-Rial, and R.~W. Heath, ``Radar aided
  beam alignment in mmwave {V2I} communications supporting antenna diversity,''
  in \emph{Prof. Inf. Theory Applic. Workshop}.\hskip 1em plus 0.5em minus
  0.4em\relax IEEE, Jun. 2016, pp. 1--7.

\bibitem{ali2019millimeter}
A.~Ali, N.~Gonz{\'a}lez-Prelcic, and A.~Ghosh, ``Millimeter wave {V2I}
  beam-training using base-station mounted radar,'' in \emph{Proc. IEEE Radar
  Conf.}\hskip 1em plus 0.5em minus 0.4em\relax IEEE, 2019, pp. 1--5.

\bibitem{liu2020radar}
F.~Liu, W.~Yuan, C.~Masouros, and J.~Yuan, ``Radar-assisted predictive
  beamforming for vehicular links: Communication served by sensing,''
  \emph{arXiv preprint arXiv:2001.09306}, 2020.

\bibitem{wymeersch20175g}
H.~Wymeersch, G.~Seco-Granados, G.~Destino, D.~Dardari, and F.~Tufvesson,
  ``{5G} mmwave positioning for vehicular networks,'' \emph{IEEE Wireless
  Commun.}, vol.~24, no.~6, pp. 80--86, Jun. 2017.

\bibitem{palacios2019hybrid}
J.~Palacios, J.~Rodriguez-Fernandez, and N.~Gonzalez-Prelcic, ``Hybrid
  precoding and combining for full-duplex millimeter wave communication,'' in
  \emph{Proc. IEEE Global Commun. Conf.}\hskip 1em plus 0.5em minus 0.4em\relax
  IEEE, 2019, pp. 1--6.

\bibitem{marzetta2016fundamentals}
T.~L. Marzetta, \emph{Fundamentals of massive {MIMO}}.\hskip 1em plus 0.5em
  minus 0.4em\relax Cambridge University Press, 2016.

\bibitem{eaves2012principles}
J.~Eaves and E.~Reedy, \emph{Principles of modern radar}.\hskip 1em plus 0.5em
  minus 0.4em\relax Springer Science \& Business Media, 2012.

\bibitem{feuerverger1977empirical}
A.~Feuerverger, R.~A. Mureika \emph{et~al.}, ``The empirical characteristic
  function and its applications,'' \emph{The Annals Stat.}, vol.~5, no.~1, pp.
  88--97, Jan. 1977.

\bibitem{ihler2005loopy}
A.~T. Ihler, W.~F. John~III, and A.~S. Willsky, ``Loopy belief propagation:
  Convergence and effects of message errors,'' \emph{J. Mach. Learn. Res.},
  vol.~6, no. May, pp. 905--936, 2005.

\bibitem{zhang2019multiple}
J.~Zhang, E.~Bj{\"o}rnson, M.~Matthaiou, D.~W.~K. Ng, H.~Yang, and D.~J. Love,
  ``Multiple antenna technologies for beyond {5G},'' \emph{arXiv preprint
  arXiv:1910.00092}, 2019.

\end{thebibliography}
\end{document}